\documentclass[%
reprint,
superscriptaddress,
 amsmath,amssymb,
 aps,
floatfix,
]{revtex4-2}
\usepackage{xcolor}
\usepackage{graphicx}
\usepackage{dcolumn}
\usepackage{bm}

\begin{document}

\preprint{APS/123-QED}

\title{Interpretable AI forecasting for numerical relativity waveforms \\ of quasi-circular, spinning, non-precessing binary black hole mergers}

\author{Asad Khan}
\affiliation{Department of Physics, University of Illinois at Urbana-Champaign, Urbana, Illinois 61801, USA}
\affiliation{Data Science and Learning Division, Argonne National Laboratory, Lemont, Illinois 60439,
USA}
\affiliation{National Center for Supercomputing Applications, University of Illinois at Urbana-Champaign, Urbana, Illinois 61801, USA}
\author{E. A. Huerta}
\affiliation{Data Science and Learning Division, Argonne National Laboratory, Lemont, Illinois 60439,
USA}
\affiliation{Department of Computer Science, University of Chicago, Chicago, Illinois 60637, USA}
\affiliation{Department of Physics, University of Illinois at Urbana-Champaign, Urbana, Illinois 61801, USA}
\author{Huihuo Zheng}
\affiliation{Leadership Computing Facility, Argonne National Laboratory, Lemont, Illinois 60439, USA}

\date{\today}

\begin{abstract}
\noindent We present a deep-learning 
artificial intelligence model that is 
capable of learning and forecasting the late-inspiral, 
merger and ringdown of numerical relativity waveforms that 
describe quasi-circular, spinning, non-precessing binary 
black hole mergers. We used the \texttt{NRHybSur3dq8} 
surrogate model to produce train, validation and 
test sets of \(\ell=|m|=2\) waveforms that cover the 
parameter space of binary black hole mergers 
with mass-ratios \(q\leq8\) and individual 
spins \(|s^z_{\{1,\,2\}}| \leq 0.8\). These waveforms 
cover the time range \(t\in[-5000\textrm{M}, 130\textrm{M}]\), 
where \(t=0M\) marks the merger event, defined as the 
maximum value of the waveform amplitude. We harnessed the 
ThetaGPU supercomputer at the Argonne Leadership Computing 
Facility to train our AI model using a training set of 
1.5 million waveforms. We used 16 NVIDIA DGX A100 nodes, 
each consisting of 8 NVIDIA A100 Tensor Core GPUs and 2 
AMD Rome CPUs, to fully train our model within 3.5 hours. 
Our findings show that artificial intelligence can accurately 
forecast the dynamical evolution of numerical 
relativity waveforms in the time range \(t\in[-100\textrm{M}, 130\textrm{M}]\). Sampling a test set of 190,000 waveforms, 
we find that the average overlap between target and 
predicted waveforms 
is $\gtrsim99\%$ over the entire parameter space under 
consideration. We also 
combined scientific visualization and accelerated computing 
to identify what components of our model take in knowledge 
from the early and 
late-time waveform evolution to accurately forecast 
the latter part of 
numerical relativity waveforms. This work aims to accelerate the 
creation of scalable, computationally efficient and 
interpretable artificial 
intelligence models for gravitational wave astrophysics.
\end{abstract}

\maketitle


\section{\label{sec:intro}Introduction}

\noindent The combination of artificial 
intelligence (AI) and 
innovative computing has led to novel, 
computationally efficient and scalable  
methodologies for gravitational wave detection~\cite{geodf:2017a,GEORGE201864,2018GN,2020arXiv200914611S,Lin:2020aps,Wang:2019zaj,Fan:2018vgw,Li:2017chi,Deighan:2020gtp,Miller:2019jtp,Krastev:2019koe,2020PhRvD.102f3015S,Dreissigacker:2020xfr,Adam:2018prd,Dreissigacker:2019edy,2020PhRvD.101f4009B,2021arXiv210810715S}, 
denoising~\cite{shen2019denoising,Wei:2019zlc,PhysRevResearch.2.033066}, parameter estimation~\cite{Shen:2019vep,Gabbard:2019rde,Chua:2019wwt,Green:2020hst,Green:2020dnx,2021arXiv210612594D}, 
rapid waveform production~\cite{Khan:2020fso,PhysRevLett.122.211101},
and early warning systems for 
multi-messenger sources~\cite{2020arXiv201203963W,Wei:2020sfz,2021PhRvD.104f2004Y}, to mention a few. 
The convergence of 
AI, distributed computing and scientific data 
infrastructure has enabled the creation of 
production scale, AI-driven frameworks for 
gravitational wave detection~\cite{KHAN2020135628,2021PhLB..81236029W,2020arXiv201208545H}. 
The fact that these advances 
have stemmed from prototypes to search for gravitational waves in advanced Laser Interferometer Gravitational Wave Observatory (LIGO) data~\cite{GEORGE201864} into production scale 
AI frameworks that process advanced LIGO data in bulk~\cite{2020arXiv201203963W,Wei:2020sfz,2021PhRvD.104f2004Y} within just five years, and that 
these methodologies have been embraced and developed 
by multiple teams around the world, furnish evidence for 
the transformational, global impact of AI and 
innovative computing in 
gravitational wave astrophysics~\cite{Nat_Rev_2019_Huerta,huerta_book,2020arXiv200503745C}.

AI has also been harnessed to learn and describe 
multi-scale and multi-physics phenomena, such as the physics 
of subgrid-scale ideal magnetohydrodynamics turbulence 
of 2D simulations of the magnetized 
Kelvin-Helmholtz instability~\cite{2020PhRvD.101h4024R}. The creation of 
AI surrogates is an active area of research that aims to 
improve the computational efficiency, scalability 
and accuracy of scientific software utilized in 
conjunction with high-performance computing (HPC) 
platforms to study and simulate complex phenomena~\cite{2019JCoPh.394...56Z,Anirudh9741}. 
It is in the spirit of this work, that researchers 
have explored the ability of AI to forecast the 
non-linear behavior of waveforms that describe the 
physics of quasi-circular, non-spinning, binary 
black hole mergers~\cite{dl_rg_lee}. 

In this study we quantify 
the ability of AI to learn and describe the 
highly dynamical, non-linear 
behaviour of numerical relativity waveforms that 
describe quasi-circular, spinning, non-precessing 
binary black hole mergers. To do this, we have 
implemented a deep-learning AI model that takes 
as input time-series waveform data that
describes the inspiral evolution, and then 
outputs time-series data that describes the 
late-inspiral, merger and ringdown of binary black 
holes that span systems with mass-ratios 
\(1\leq q \leq 8\), and individual 
spins \(s^z_{{1,2}}\in[-0.8, 0.8]\). To make apparent 
the size and complexity of this problem, the astute 
reader may notice that the amount of training data to 
address this problem in the context of non-spinning, 
quasi-circular binary black hole mergers is of 
order \(\sim1.2\times10^{4}\)~\cite{dl_rg_lee}. 
In stark contrast, addressing this problem in the 
context of quasi-circular, spinning, non-precessing binary 
black hole mergers requires a training dataset that 
contains over \(\sim1.5\times10^{6}\) modeled 
waveforms to densely sample this high dimensional 
signal manifold. This amount of data is needed to 
capture the rich dynamics imprinted in the waveforms 
that describe these astrophysical systems. The strategy 
we have followed to tackle this computational grand 
challenge consists of combining AI and HPC to 
reduce time-to-insight, and to incorporate a number 
of methodologies to create our Transformer based AI model, including 
positional encoding, 
multi-head self-attention, multi-head cross attention, 
layer normalization, and residual connections. 

Furthermore, we acknowledge the importance 
of going beyond innovative algorithm design, 
and the confluence of AI and HPC to address these 
types of computational challenges. There is a pressing 
need to understand how 
AI models abstract knowledge 
from data and make predictions. Thus, we also showcase 
the use of scientific visualization and 
HPC to \textit{interpret} and \textit{understand} 
how various components of our AI model work together 
to make accurate predictions. Throughout this paper 
we use geometric units in which \(G=c=1\). In this 
convention, \(M\) sets the length scale of the 
scale invariant black hole simulations, and 
corresponds to the total mass of the spacetime 
simulated. For instance, 
\(M = 1M_{\odot}=4.93\times10^{-6}\textrm{ s}\) or
\(M = 1M_{\odot}=1.48\textrm{ km}\). 
In this article we 
use \(M\) to describe time.

This article is organized as follows. 
Section~\ref{sec:method} describes the datasets, 
neural network architecture and optimization methods
used to create our AI model. We present and discuss 
our results in Section~\ref{sec:res}. This section 
includes a detailed study of the forecasting 
capabilities of our 
AI model, as well as interpretability studies. 
Finally, we summarize our findings and outline 
future work in Section~\ref{sec:end}.

\section{Methods}
\label{sec:method}

\noindent Here we describe the waveform datasets 
used for this study, the key components of our AI 
model, and the approaches followed to train 
and optimize it.

\subsection{Dataset}
\label{sec:data}

We consider inspiral-merger-ringdown waveforms 
that describe quasi-circular, spinning, 
non-precessing binary black hole mergers. We 
have produced training, test and validation waveform 
sets with the surrogate model 
\texttt{NRHybSur3dq8} \cite{PhysRevD.99.064045}. 
Since the surrogate \texttt{NRHybSur3dq8} is trained 
with 104 numerical relativity waveforms in the 
parameter range \(q \leq 8\) and \(|s^z_i| \leq 0.8\), 
we restrict our datasets to lie within the same 
parameter span. Throughout this paper we use a geometric
unit system in which \(G=c=1\). 

\begin{figure*}[htp]
\centerline{
\includegraphics[width=.53\linewidth]{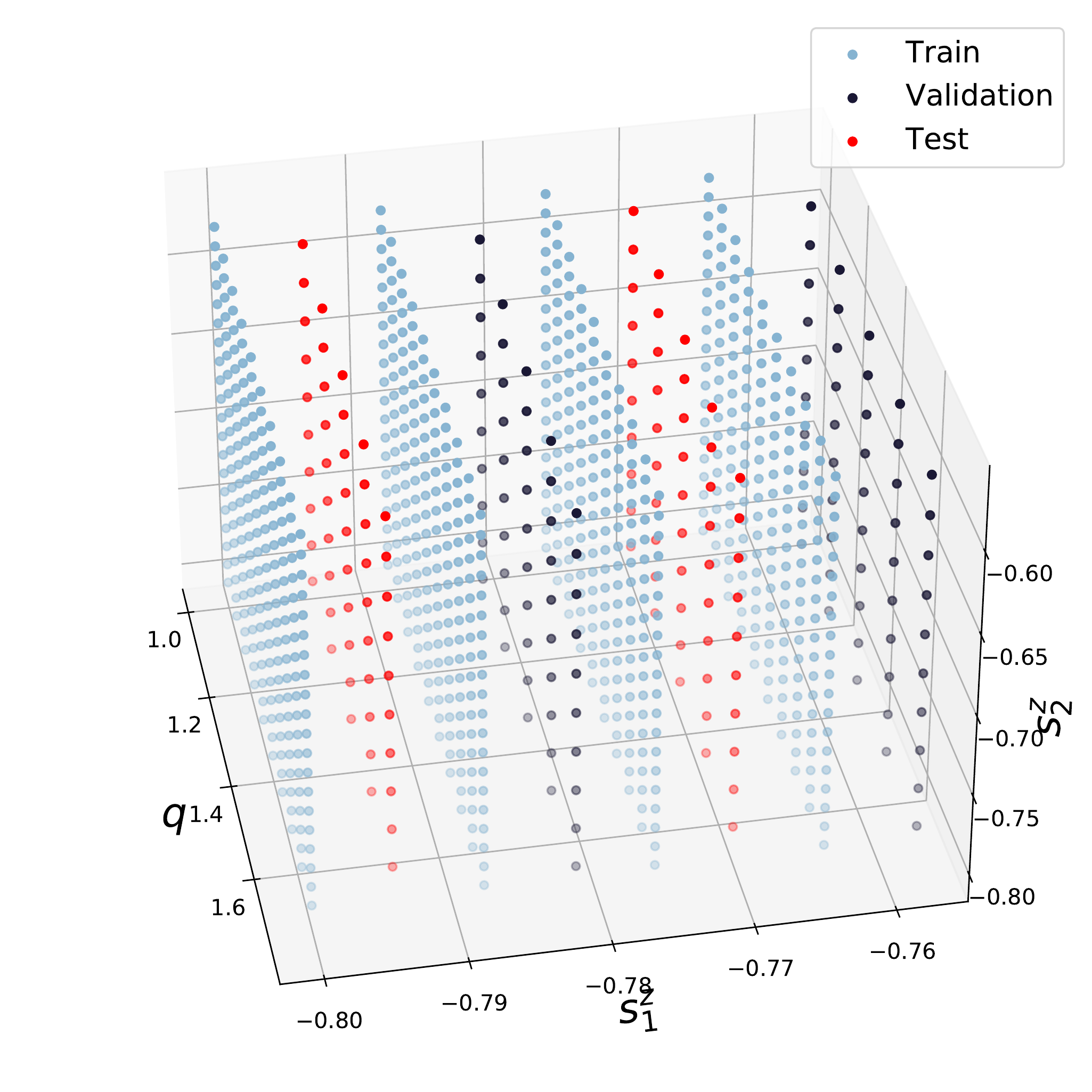}
\includegraphics[width=.47\linewidth]{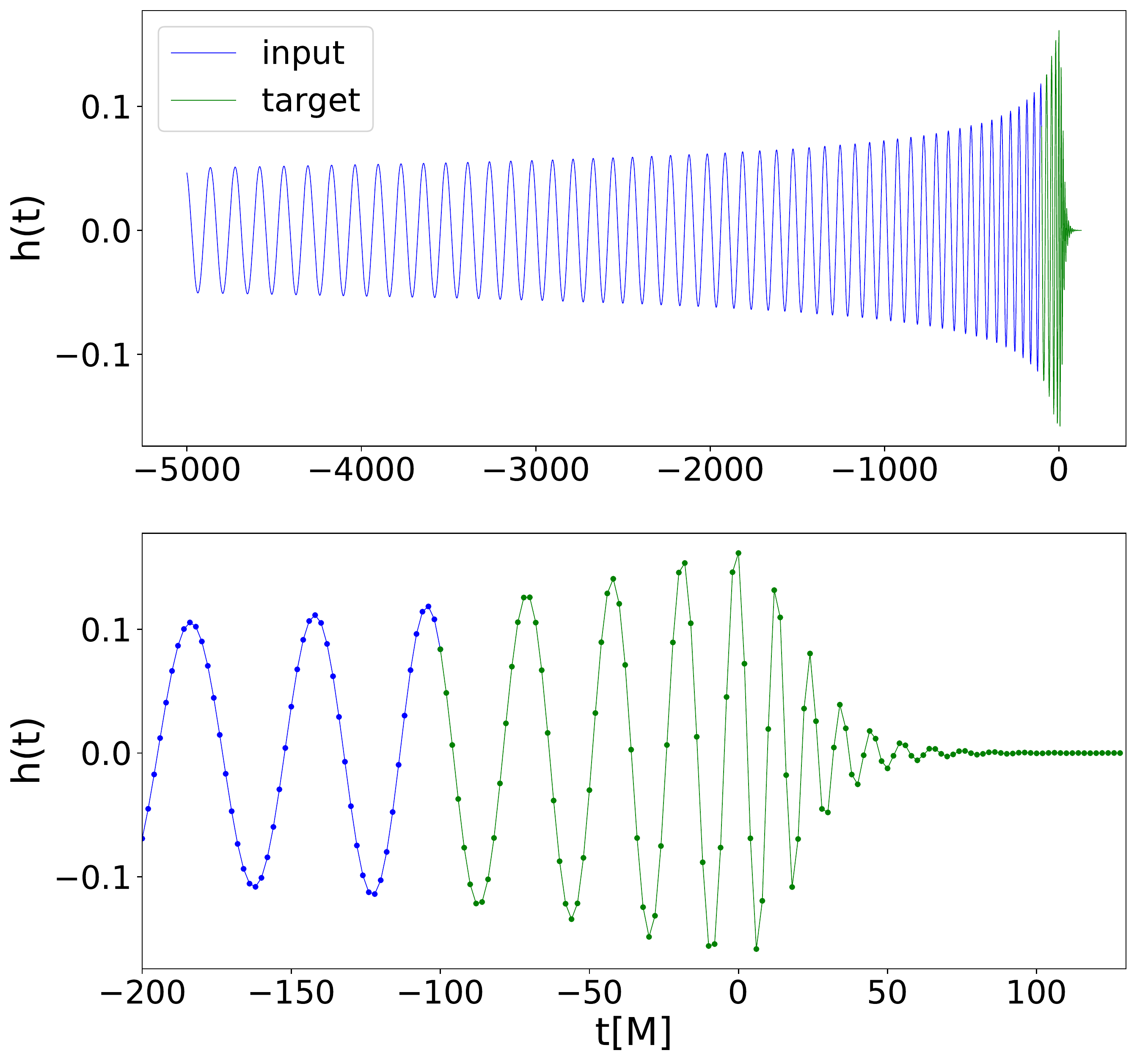}
}
\caption{\textbf{Left panel} Training, validation and 
test sets for the binary black hole 3-D signal manifold 
\(1 \leq q \leq 8\) and \(s^z_{\{1,2\}}\in[-0.8, 0.8]\). 
1.5M waveforms are used for the training set, and 
190,00 waveforms for the test and validation sets. The 
sampling shown in this 3-D representation for 
\(q \in [1, 1.8)\) is mirrored throughout the parameter 
space under consideration. \textbf{Top-right panel} Sample 
waveform for a binary black hole with parameters 
\(\{q, s^z_1, s^z_2\}=\{6.8,0.718,0.718\}\). Signals 
span the time window \(t\in[-5000M, 130M]\) sampled with
a time step \(\Delta t = 2\,\textrm{M}\). \textbf{Bottom-right panel} Input data to our AI model spans the time 
window \(t\leq -100M\), whereas \(t\geq -100M\) 
represents the target time-series output.}
\label{fig:data}
\end{figure*}

We use \(\ell=|m|=2\) waveforms for this study that 
cover the time span \(t\in[-5,000\,\textrm{M}, 
130\,\textrm{M}]\) with the merger (amplitude peak of 
the signal) occurring at \(t=0M\). To accurately capture 
the dynamics of the waveform we sample it with a time 
step \(\Delta t = 2\,\textrm{M}\). We split each 
waveform into two segments, namely, the input consisting
of the early inspiral phase covering the time 
span \(t\in[-5,000\,\textrm{M}, -100\,\textrm{M}]\), 
and the target consisting of late-inspiral, merger 
and ringdown covering the time span 
\(t\in[-100\,\textrm{M}, 130\,\textrm{M}]\). 
We then train an AI model to forecast the target 
waveform segment when fed with the input waveform 
segment. An example waveform with the input and target segments is shown in Figure~\ref{fig:data}.

The training set consists of $\sim1.5$ million 
 waveforms generated by sampling the mass-ratio 
 $q\in[1,8]$ in steps of \(\Delta q = 0.08\), and 
 the individual spins  $s^z_i\in[-0.8, 0.8]$ in steps 
 of \(\Delta s^z_i = 0.012\). The validation and test 
 sets consist of $\sim 190,000$ waveforms each, and 
 are generated by alternately sampling the 
 intermediate values, i.e., by sampling $q$ and $s^z_i$ 
 in steps of $0.16$ and $0.024$ to lie between training 
 set values. We show a small slice of the parameter 
 space to illustrate this sampling in 
 Figure~\ref{fig:data}.

\subsection{Neural network architecture}
\label{sec:architecture}

The neural network we use for numerical relativity 
waveform forecasting is a slightly modified version of 
the \texttt{Transformer} model, originally proposed 
in the context of Natural Language 
Processing (NLP)~\cite{2017arXiv170603762V}. The 
fundamental operation in the \texttt{Transformer} model 
is the multi-head scaled dot-product 
\texttt{Attention} mechanism. \texttt{Attention} can 
be thought of as a mapping between two sets; each 
element of the output set is a weighted average of 
all elements in the input set, where the weights are 
assigned according to some scoring function. This helps 
with context aware memorization of long sequences. 
We briefly discuss the various components of 
the \texttt{Transformer} model below.

\subsubsection{Scaled dot-product attention}
Consider a set of n input vectors 
$\{x_1, x_2, x_3, ..., x_n\}$ and a set of \(t\) output 
vectors $\{h_1, h_2, h_3, ..., h_t\}$ in $\mathbb{R}^d$.
Then according to scaled dot product attention, 
the outputs are computed as follows:

\begin{equation}
\label{eq1}
    h_i = \sum_j w_{ij} v_j\,,
\end{equation}

\noindent where,

\begin{align}
\label{eq2}
    w_{ij} &= \texttt{softmax}\left( \frac{q_i^T k_j}{\sqrt{d}} \right)\,,\\ 
    \label{eq3}
    q_i &= W_q x_i\,,\\ 
    \label{eq4}
    k_i &= W_k x_i\,,\\ 
    \label{eq5}
    v_i &= W_v x_i\,,
\end{align}

\noindent where $W_q$, $W_k$ and $W_v$ are three 
learnable weight matrices and each of the three 
vectors $q_i$, $k_i$, $v_i$ (referred to as queries, keys and values) are linear transformations of the specific 
input $x_i$.

\subsubsection{Self and cross attention}

Self attention refers to applying the attention 
mechanism to relate different elements of a single set, 
i.e., queries, keys and values all correspond to the 
linear transformations of the same set of vectors 
$\{x_i\}$ as above. However, in cross attention the 
queries can come from a different set of 
vectors $\{y_i\}$, i.e., $q_i = W_q y_i$. 

In our case, the set 
$\{x_1, x_2, x_3, ..., x_n\}$ corresponds to the 
input waveform segment and the set 
$\{y_1, y_2, y_3, ..., y_t\}$ corresponds to the 
target waveform segment. 
These are shown in blue and green respectively in the right panels of Figure~\ref{fig:data}.

\subsubsection{Multi-head Attention}

Multi-head attention simply refers to applying the 
attention operation several times in parallel 
to independently projected queries, keys and values, 
i.e., for \(n\) heads we would have \(n\) sets of 
the three matrices; $W_q^i$, $W_k^i$ and 
$W_v^i$, $i \in \{1,2,3,..,n\}$. 
To do this efficiently, the multi-head attention module 
first splits the input vector $x_i$ into \(n\) smaller 
chunks, and then computes the attention scores over each
of the \(n\) subspaces in parallel. 

\subsubsection{Positional encoding}

\begin{widetext}
\begin{figure*}[htp]
\centerline{
\includegraphics[width=\linewidth]{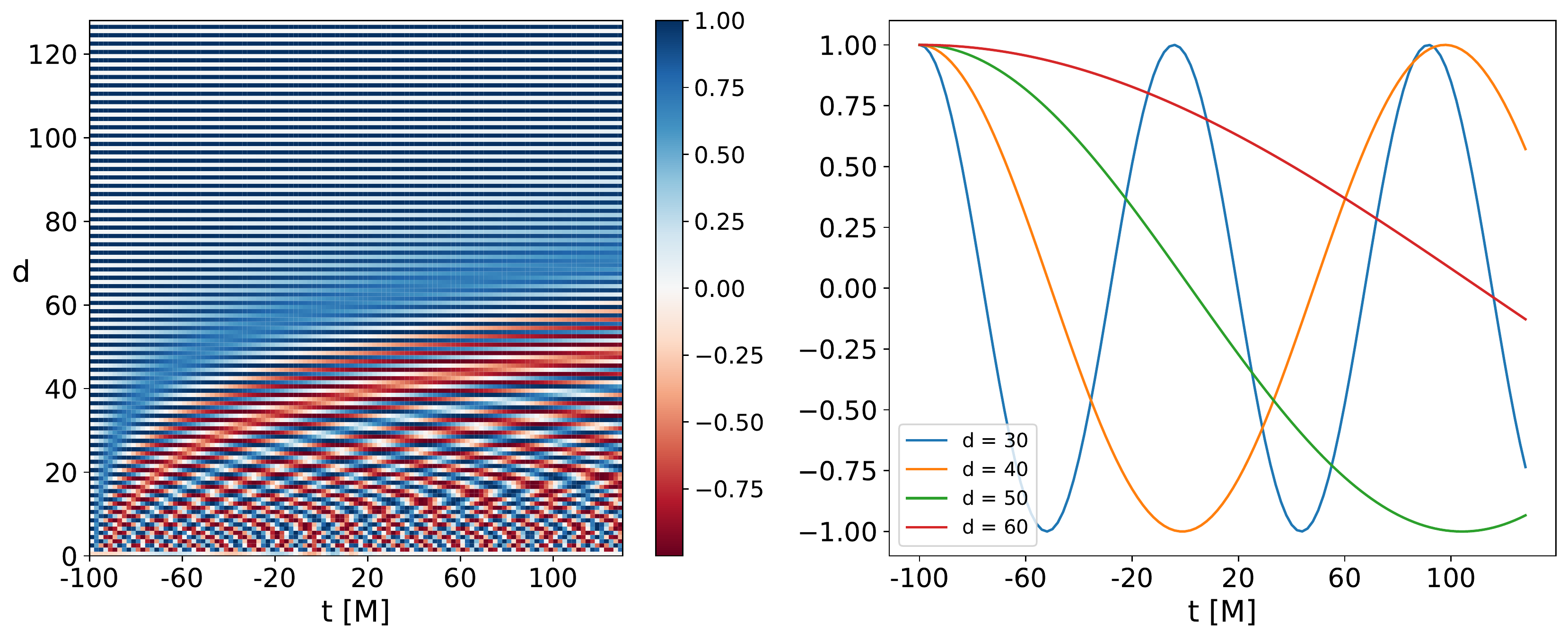}
} 
\caption{\textbf{Left panel} Heatmap of the evaluation 
of the positional encodings---
see Eqs.\eqref{eq6}-\eqref{eq7}. 
These real-valued vectors are computed at each 
timestamp of the target waveform 
\(t\in[-100\textrm{M}, 130\textrm{M}]\)---shown in 
the x-axis, for each dimension \(d\)---shown in 
the y-axis. \textbf{Right panel} Sample of encodings 
evaluated at several timestamps and dimensions. The encoding at 
each dimension \(d\) is a sinusoid of a different 
frequency.}
\label{fig:pe}
\end{figure*}
\end{widetext}

In our case, the inputs and output waveform segments 
are not sets but ordered time-series sequences. However, 
we can see from Equation~\ref{eq1} that attention 
mechanism is permutation equivariant, i.e., it 
ignores the sequential nature of the input. In order 
to make the model sensitive to the sequential ordering 
of the data, we inject information about the 
absolute positioning of the time-steps in the form 
of Positional Encoding (\texttt{PE}), i.e., some 
fixed function 
$f : \mathbb{N} \to \mathbb{R}^d$ to map the positions 
to real valued vectors. Following the 
original \texttt{Transformer} paper, we compute 
the positional encodings as follows:

\begin{align}
\label{eq6}
    \texttt{PE}(p,2i) &= \sin(p/10000^{2i/d})\,, \\
\label{eq7}    
    \texttt{PE}(p,2i+1) &= \cos(p/10000^{2i/d})\,,
\end{align}

\noindent where $p$ is the position and $i$ is the 
dimension. A sample encoding for $d=128$, used for the 
actual analysis conducted in this paper, is shown 
in Figure~\ref{fig:pe}. It is worth mentioning that
the dimension $d$ is a hyper-parameter and has to
be tuned for optimal performance. 

At the fundamental level, the input to our 
model is $1$ dimensional (a 1D wave).
However, we transform this data from rank-1 to rank-2, 
i.e., from a sequence of real numbers of
amplitude values $(h_1, h_2, …, h_n)$ to a sequence of 
$d+1$ dimensional vectors 
$(\mathbf{v_1}, \mathbf{v_2}, …, \mathbf{v_n})$, where each 
$\mathbf{v_i} = [ h_i, \texttt{PE}(i, 1), \texttt{PE}(i, 2), 
\texttt{PE}(i, 3), ... , \texttt{PE}(i, d)]$, 
and $\texttt{PE}(i, n)$ is given by equations~\ref{eq6} 
and~\ref{eq7}.

We do this because we want 
the model to be aware of the 
time-stamp of each amplitude value.
One could in principle do this by inputting into 
the model a tuple $(h_i, t_i)$ instead of just the
sequence of amplitude values $(h_i,)$. 
However, positional encodings in the manner described
above have historically worked much better.

\subsubsection{Encoder and decoder modules}

\noindent The \texttt{Transformer} model consists 
of an encoder 
module and a decoder module. The encoder takes in 
an input sequence $\{x_1, x_2,$ $x_3, ..., x_n\}$, passes 
it through a multi-head self-attention layer and 
a position-wise fully connected feed-forward network, 
mapping it to an attention based latent vector 
representation 
$\{h_1, h_2, h_3, ..., h_n\}$. This 
latent representation is then passed to the decoder module, 
which outputs the desired target sequence 
$\{y_1, y_2, y_3, ..., y_t\}$. At each time-step 
$t=i$ when the decoder is predicting $y_i$, it passes 
the thus-far generated output sequence 
$\{y_1, y_2, y_3, ..., y_{i-1}\}$ through a 
multi-head self-attention layer and the latent 
vector representation $\{h_1, h_2, h_3, ..., h_n\}$ 
through a multi-head cross attention layer. The 
two are added together and passed through a 
position-wise fully connected feed-forward network 
and a final 1D convolutional layer to generate the 
next time-step of the output sequence $y_{i}$ in 
an autoregressive fashion. 

\begin{figure*}[htp]
\centerline{
\includegraphics[width=.6\linewidth]{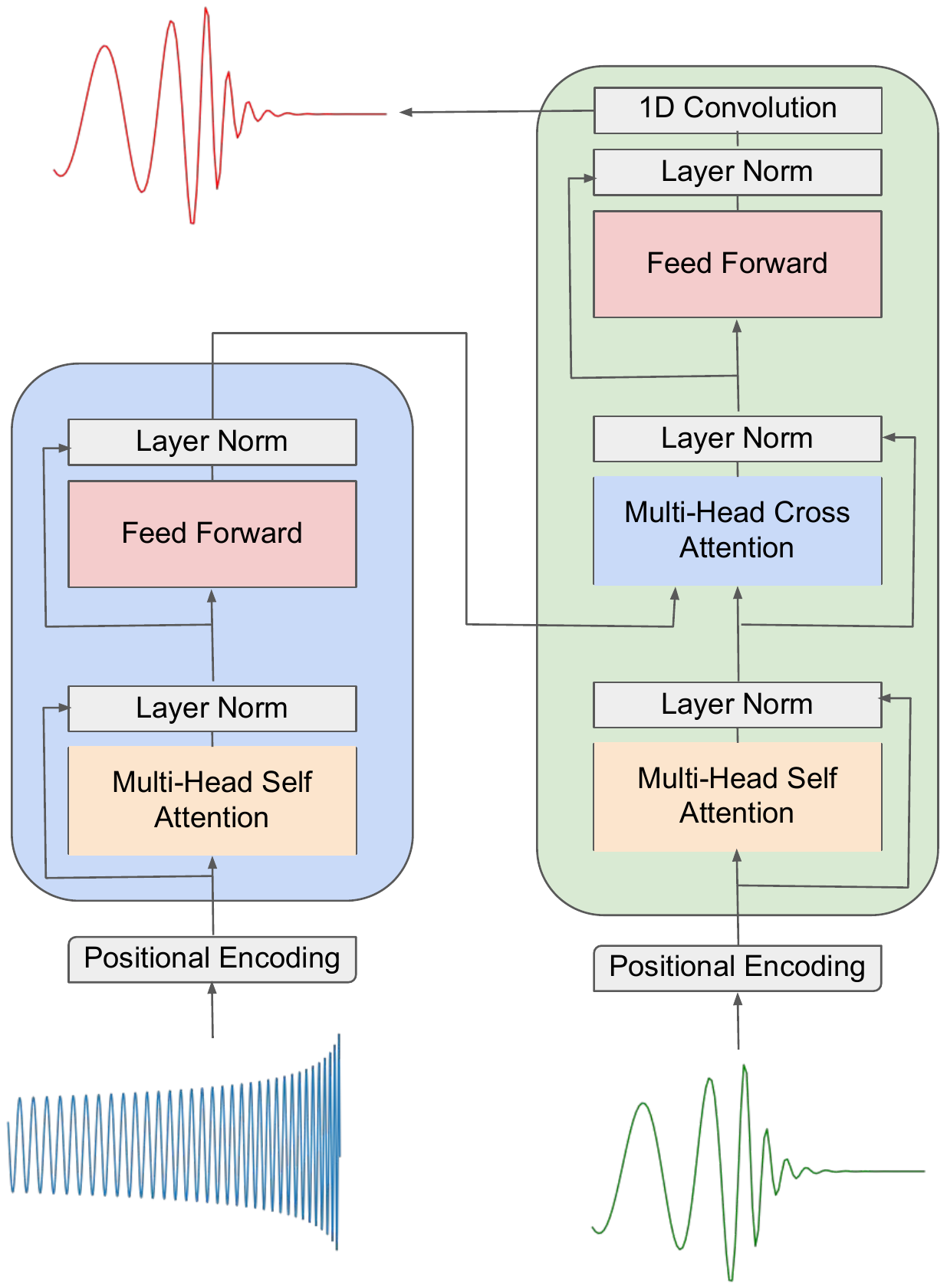}
} 
\caption{\textbf{Model architecture} Schematic 
representation of our AI model. During training we 
provide two input waveforms, 
namely, a pre-merger 
waveform that spans the time 
range \(t\leq -100\textrm{M}\)---shown at the bottom 
left of the diagram; and a time-shift version of the 
target waveform that spans the time 
range \(t\in[-101\textrm{M}, 129\textrm{M}]\)---shown 
at the bottom right of the diagram. The output of 
this AI model, the target waveform that spans the 
range \(t\in[-100\textrm{M}, 130\textrm{M}]\), is 
shown at the top left of this diagram. At inference, 
we provide an input waveform---as indicated in the 
bottom left of the diagram. The model then outputs 
time samples up to time \(i\), which are then 
passed as input---as shown in the bottom right panel 
of the figure---so that the model produces the following 
time samples up to time \(i+1\). The final output is 
a waveform that covers the range \(t\in[-100\textrm{M}, 130\textrm{M}]\).}
\label{fig:model}
\end{figure*}

Both the encoder and decoder modules also make use 
of layer normalization and residual connections. We 
refer the reader for a more in depth discussion of 
the \texttt{Transformer} model to the original 
paper~\cite{2017arXiv170603762V}. We summarize the architecture for our model in Figure~\ref{fig:model}.

\subsection{Training and optimization}
\label{sec:trainopt}

As mentioned above, we first divide the 
waveforms into input segments corresponding 
to \(t\in[-5000\,\textrm{M}, -100\,\textrm{M}]\), and 
target segments corresponding to 
\(t\in[-100\,\textrm{M}, 130\,\textrm{M}]\). We 
then concatenate both segments with their respective 
fixed positional encodings. In our experiments, we 
trained two models; one on only the plus polarization waveforms and 
another on a dataset composed of equal number of plus and cross 
polarization waveforms. However, we did not find a significant 
difference in the performance between these two models, i.e. the model 
trained on only plus polarizations was just as good at generalizing to 
the cross polarizations, as the model that was trained on both. 
Consequently, in this paper we report results for the model that was 
trained only on the plus polarization, but during inference it is used 
to predict both plus and cross polarization.
During training time we 
employ the \texttt{Teacher Forcing} methodology, i.e., 
we pass the input segment through the encoder, 
and a one step time-shifted version of the target 
to the decoder. This means that true output is fed 
to the decoder for the next time step prediction 
regardless of the predicted value at the current 
time-step, which helps the model converge faster. 
A visual exposition of this methodology 
is presented in Appendix~\ref{sec:ap1}.

We use mean-squared error (MSE) between the predicted 
and the target series as the loss function, and use 
Adam optimizer with $\beta_1 = 0.9, \beta_2 = 0.999, 
\epsilon = 1e-07$ and learning\_rate $ = 0.001$. 
During training we also monitor the loss on the 
validation set to prevent over-fitting and to 
dynamically reduce the learning rate whenever the 
loss hits a plateau.

We trained our AI model using 16 NVIDIA DGX A100 nodes 
at the Argonne Leadership Computing Facility. Each node 
comprises of eight NVIDIA A100 Tensor Core GPUs 
and two AMD Rome CPUs that provide 320 gigabytes of GPU 
memory. We used a batch-size of 8 and trained the model 
for a total of 53 epochs, reaching convergence in 3.5 hours.

\section{Results}
\label{sec:res}

\begin{widetext}
\begin{figure*}[htp]
\centerline{
\includegraphics[width=0.33\linewidth]{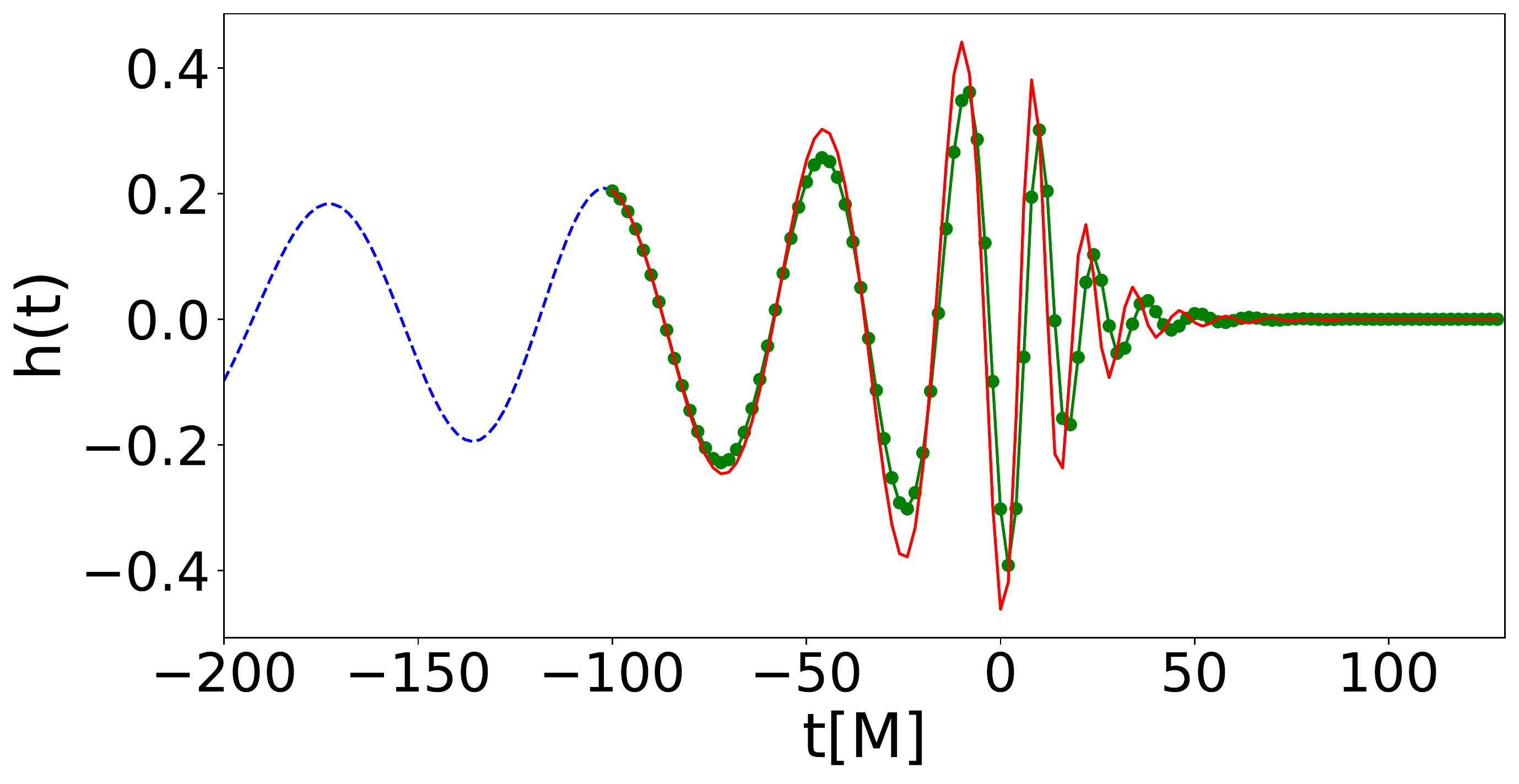}
\includegraphics[width=0.33\linewidth]{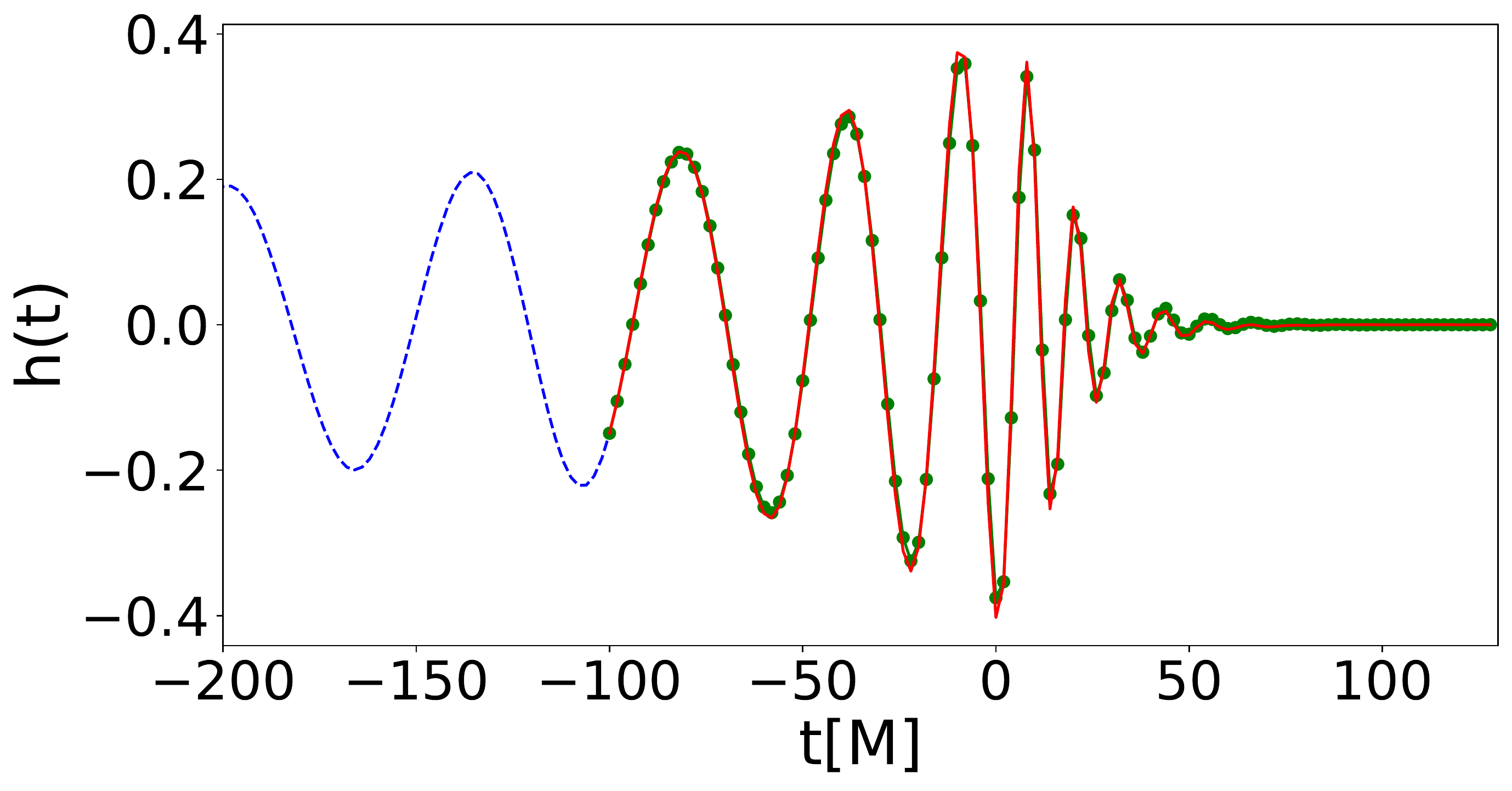}
\includegraphics[width=0.33\linewidth]{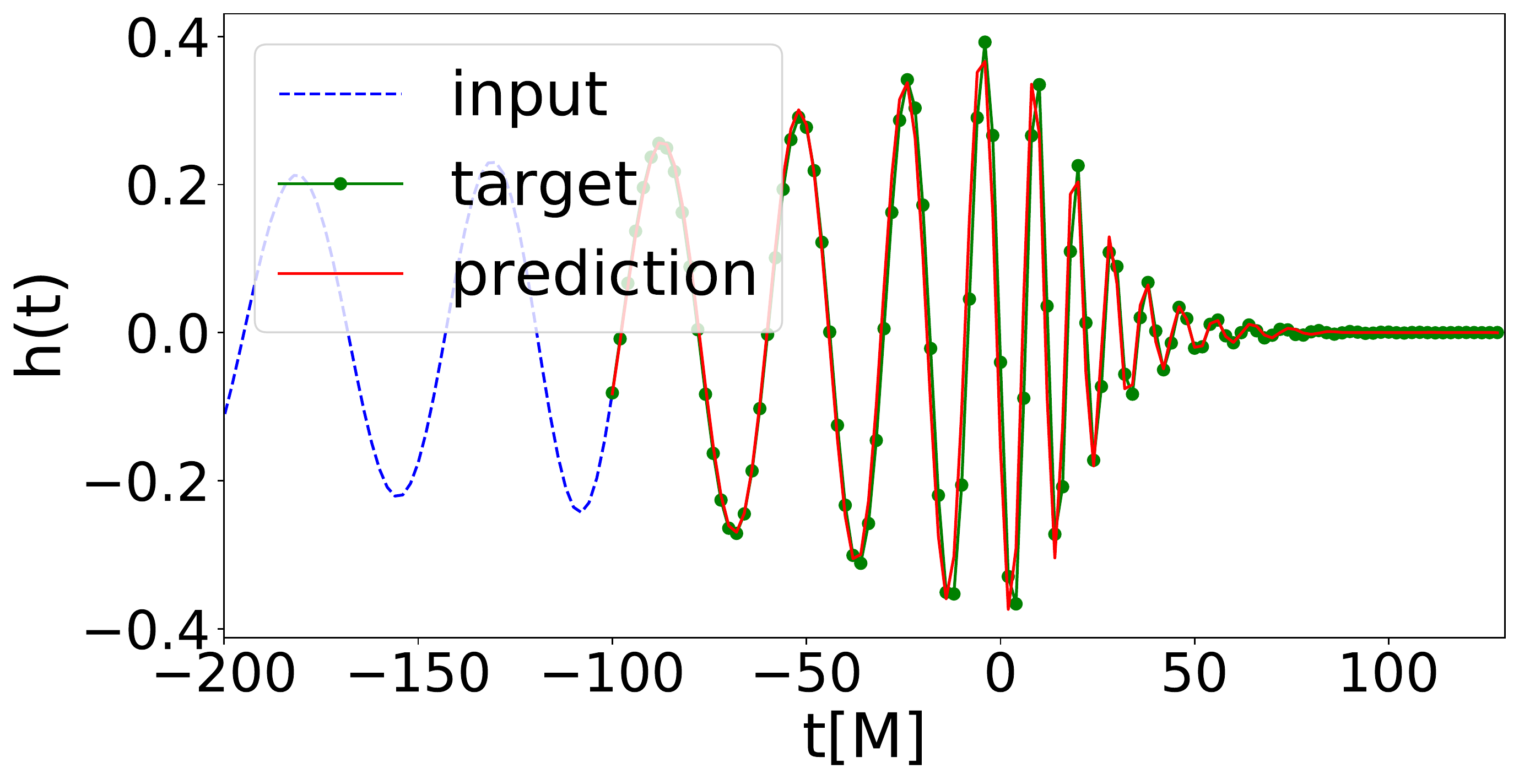}
} 
\centerline{
\includegraphics[width=0.33\linewidth]{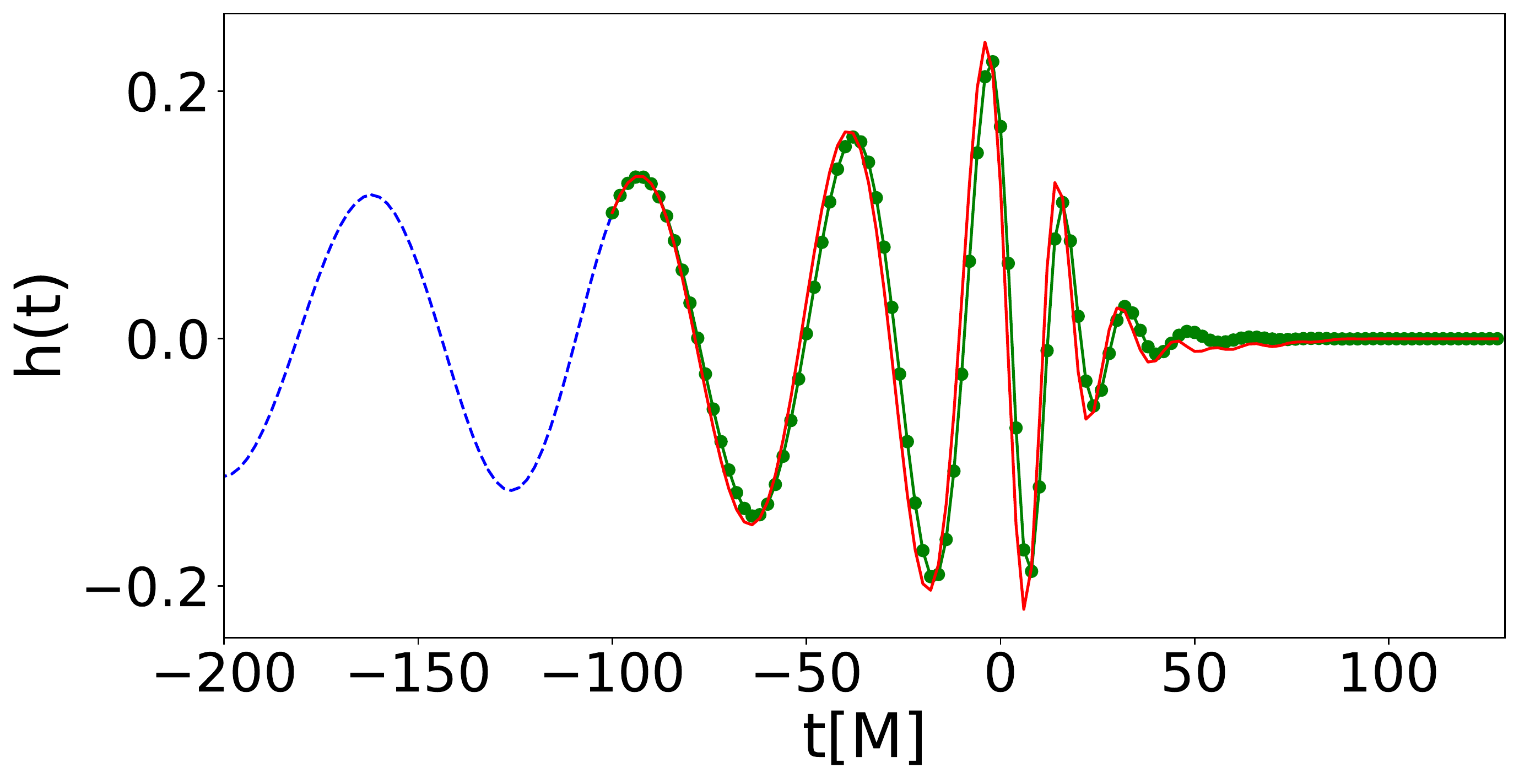}
\includegraphics[width=0.33\linewidth]{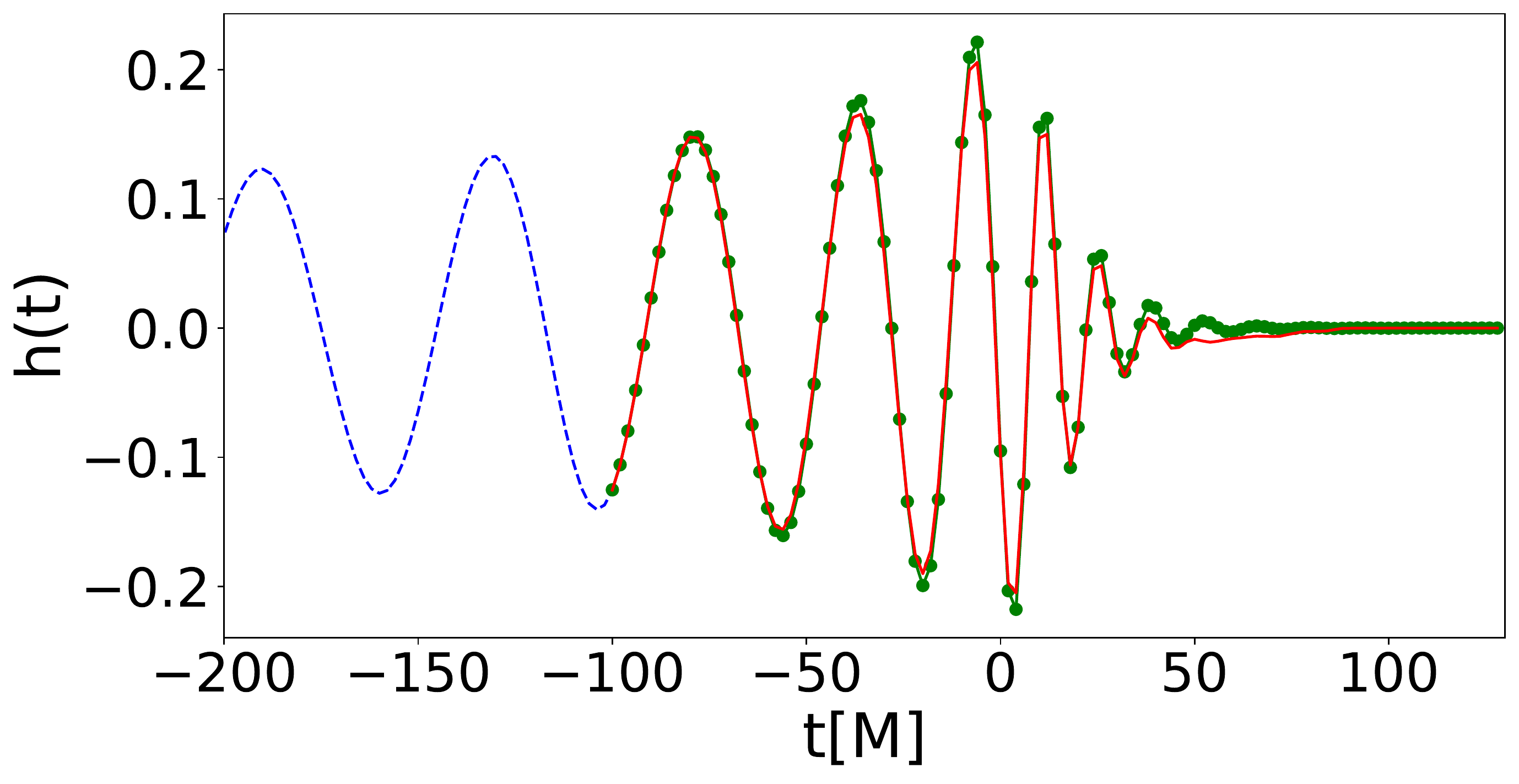}
\includegraphics[width=0.33\linewidth]{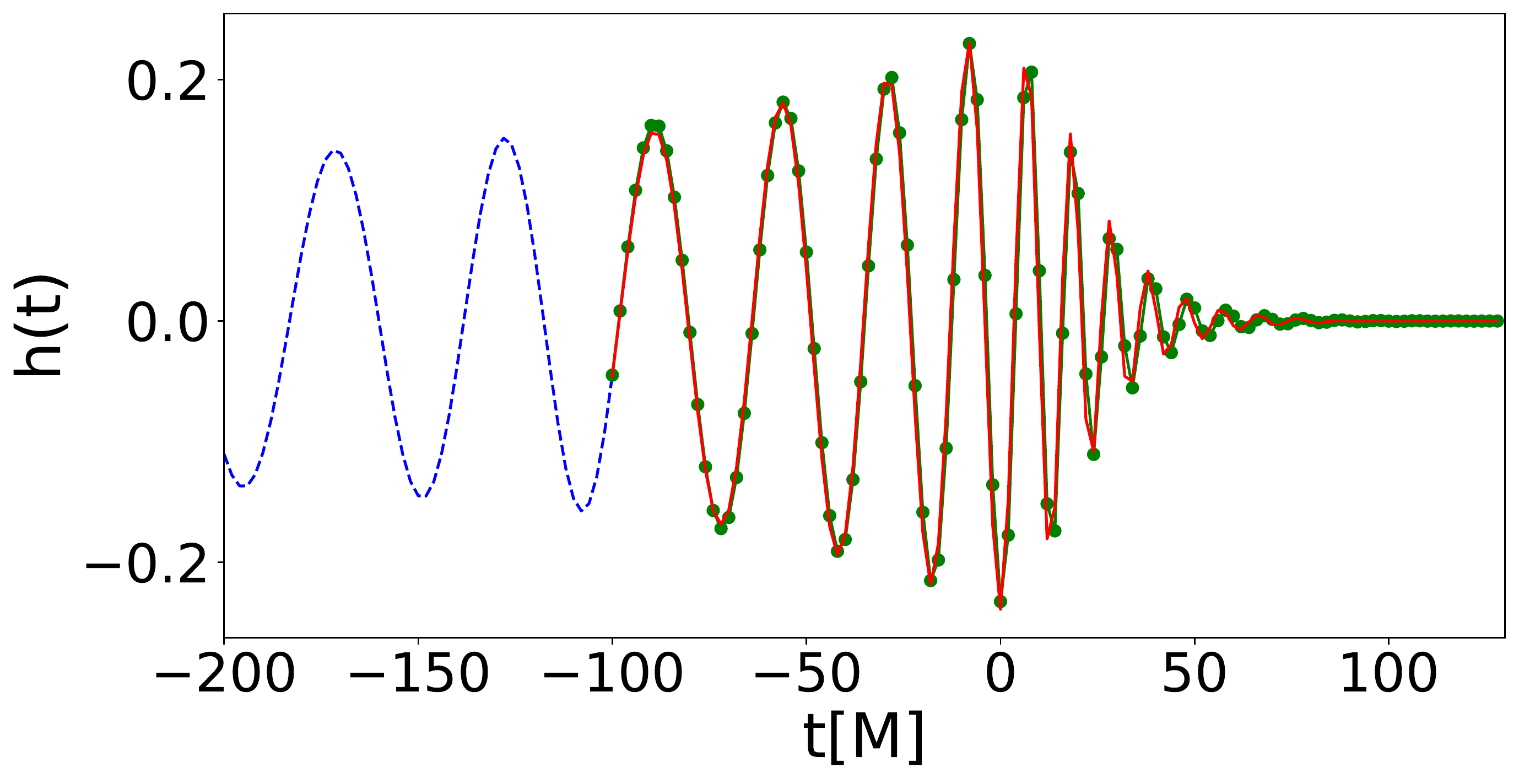}
} 
\centerline{
\includegraphics[width=0.33\linewidth]{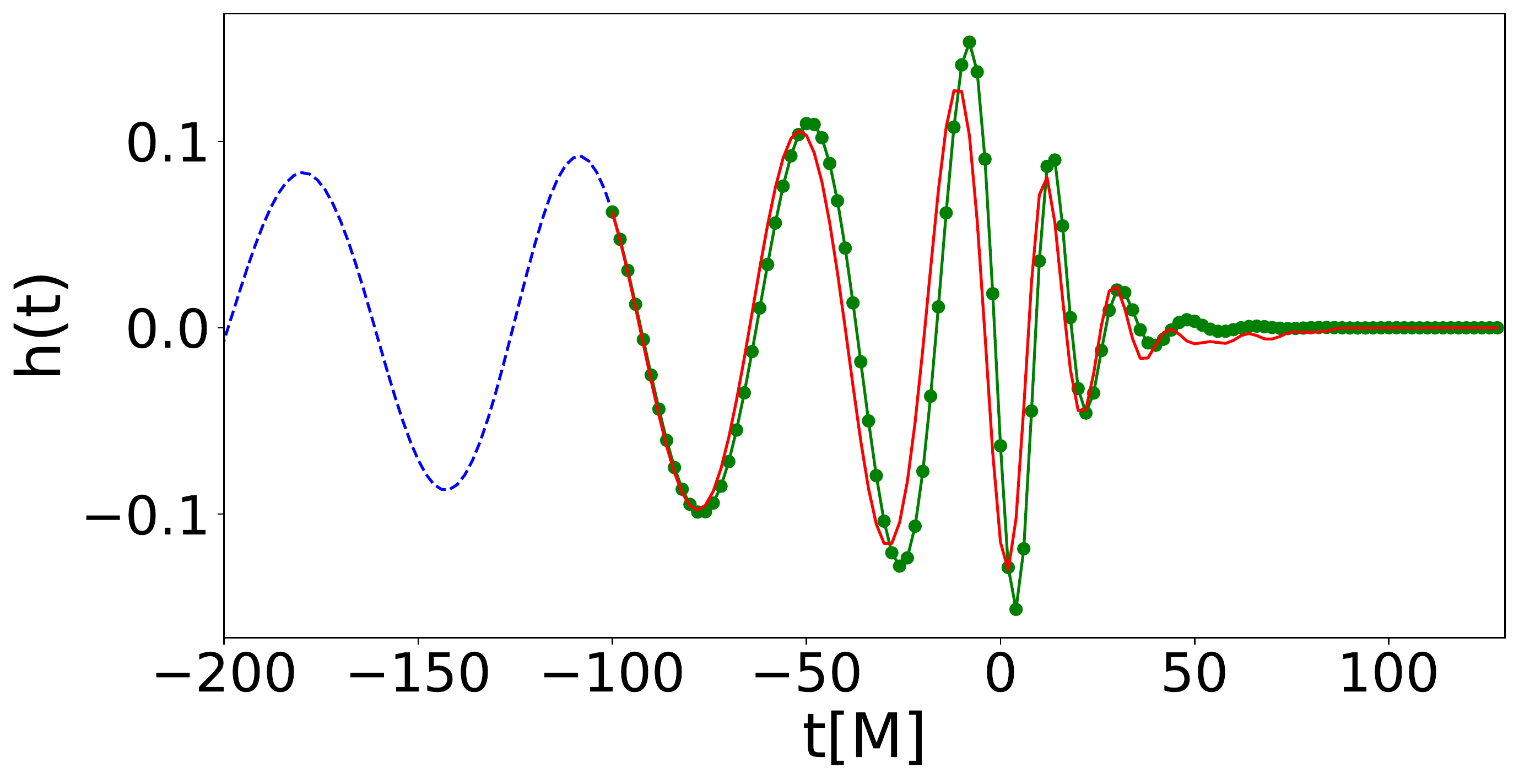}
\includegraphics[width=0.33\linewidth]{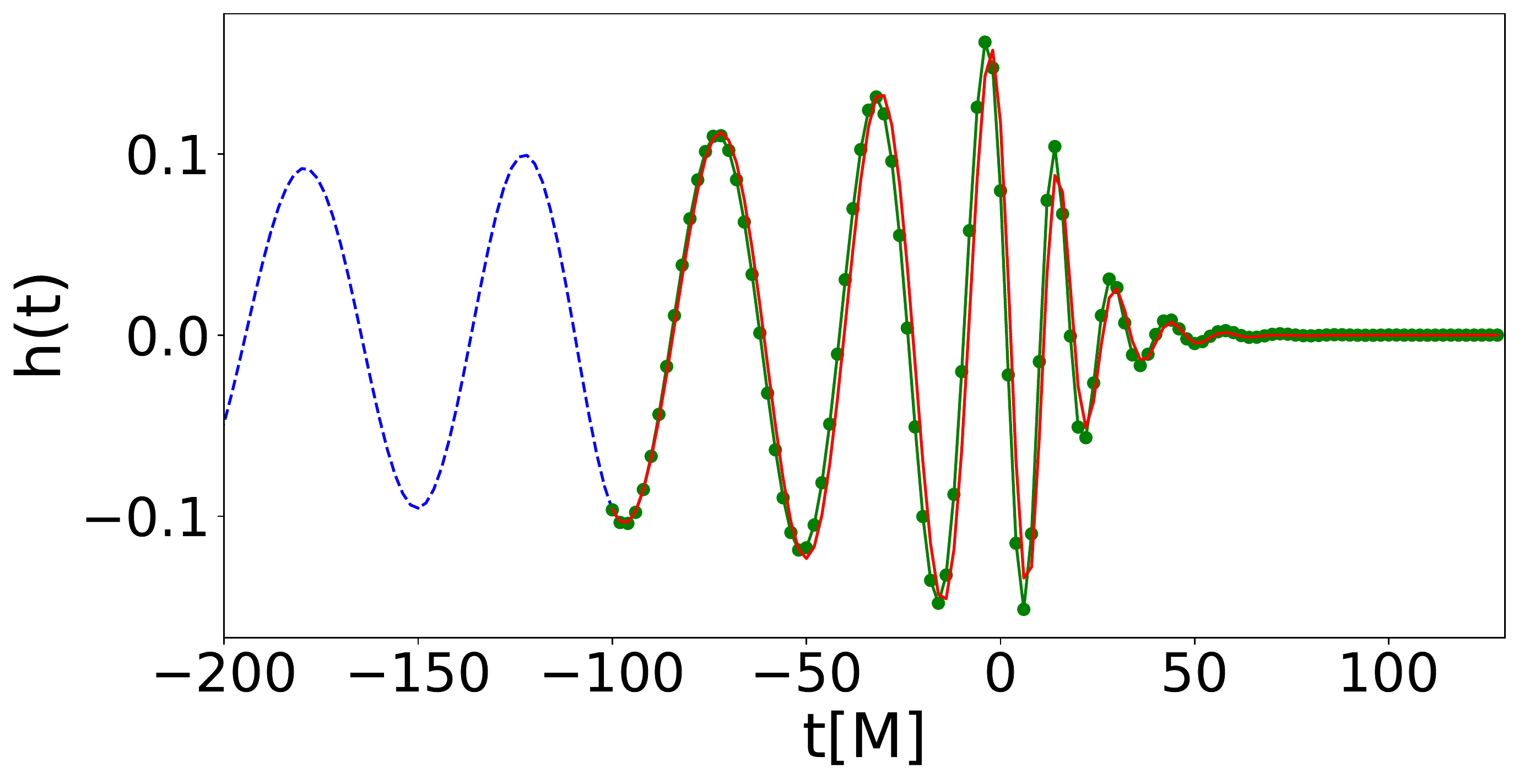}
\includegraphics[width=0.33\linewidth]{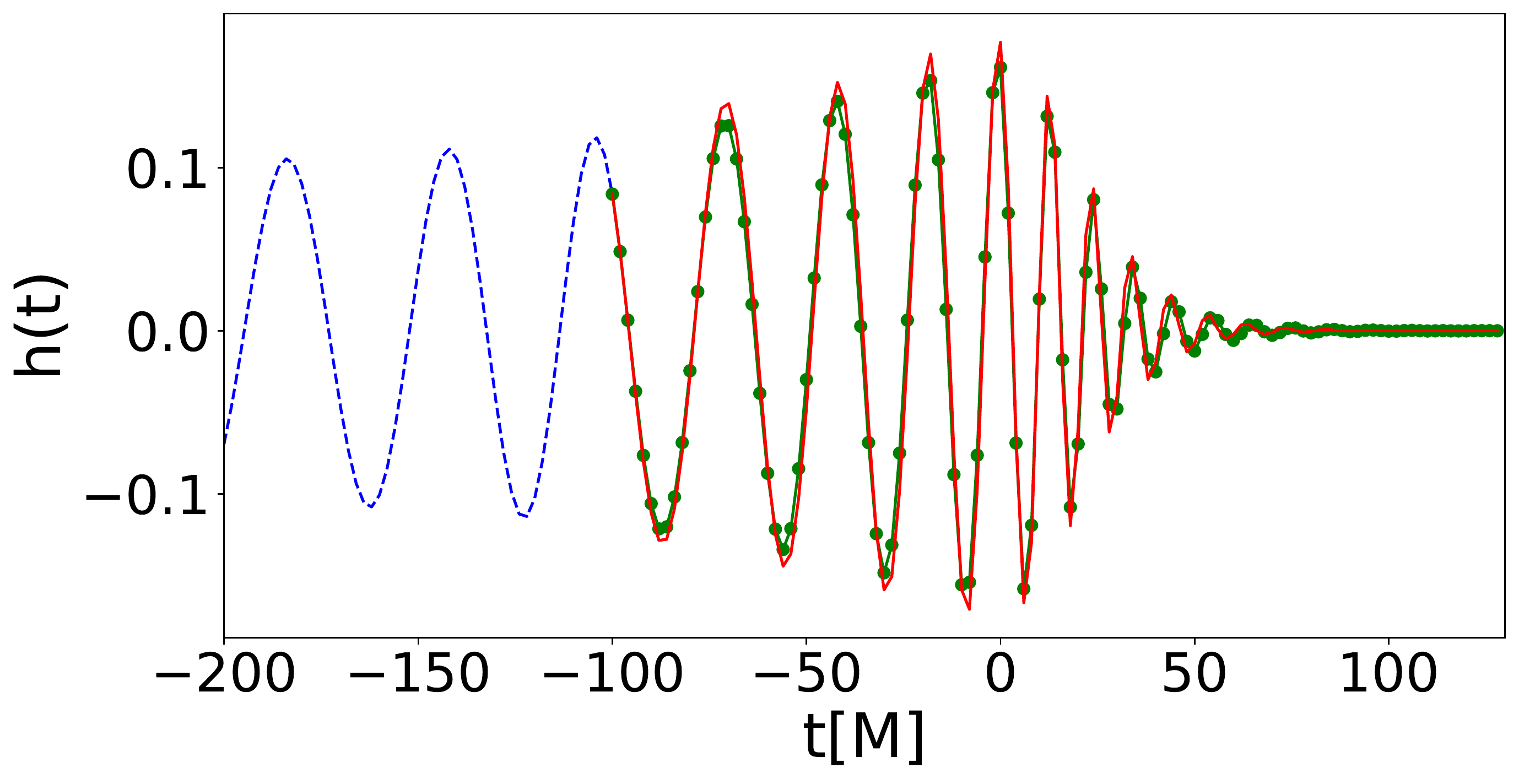}
}
\caption{\textbf{Gallery of results} Sample input, 
target and predicted waveforms for binary black holes 
with mass-ratios \(q=\{1.04, 4.24, 6.80\}\), from top 
to bottom; and spins \(s^z_1=s^z_2= 
\{-0.7,0.0, 0.7\}\), from left to right. Notice the 
impact of individual spins 
in the dynamics of the systems, encompassing rapid 
(left column) and delayed plunges 
(right column). The 
model predicts the waveform evolution in 
the range \(-100\textrm{M} \leq t \leq 130\textrm{M}\).}
\label{fig:examples}
\end{figure*}
\end{widetext}

During inference, we only feed the input 
segment to the model and let it recover the 
full target sequence autoregressively, i.e., to make 
the prediction at time-step $t=i$, the decoder module 
is fed its own prediction from the previous 
time-step $t=i-1$. Our AI model outputs both the plus 
and cross polarizations. We show a representative sample 
of target and predicted waveforms 
in Figure~\ref{fig:examples}. We have selected these 
cases to provide a visual representation of the rich 
dynamics captured by our AI model, encompassing 
rapid plunges represented by black hole binaries whose 
components have negative spins (left column); non-spinning 
binary black holes (mid column); and systems that, 
on account of having binary components with positive 
spins and thus more angular momentum, complete 
more waveform cycles before plunge (right column). 

\begin{figure*}[ht]
\centerline{
\includegraphics[width=0.33\linewidth]{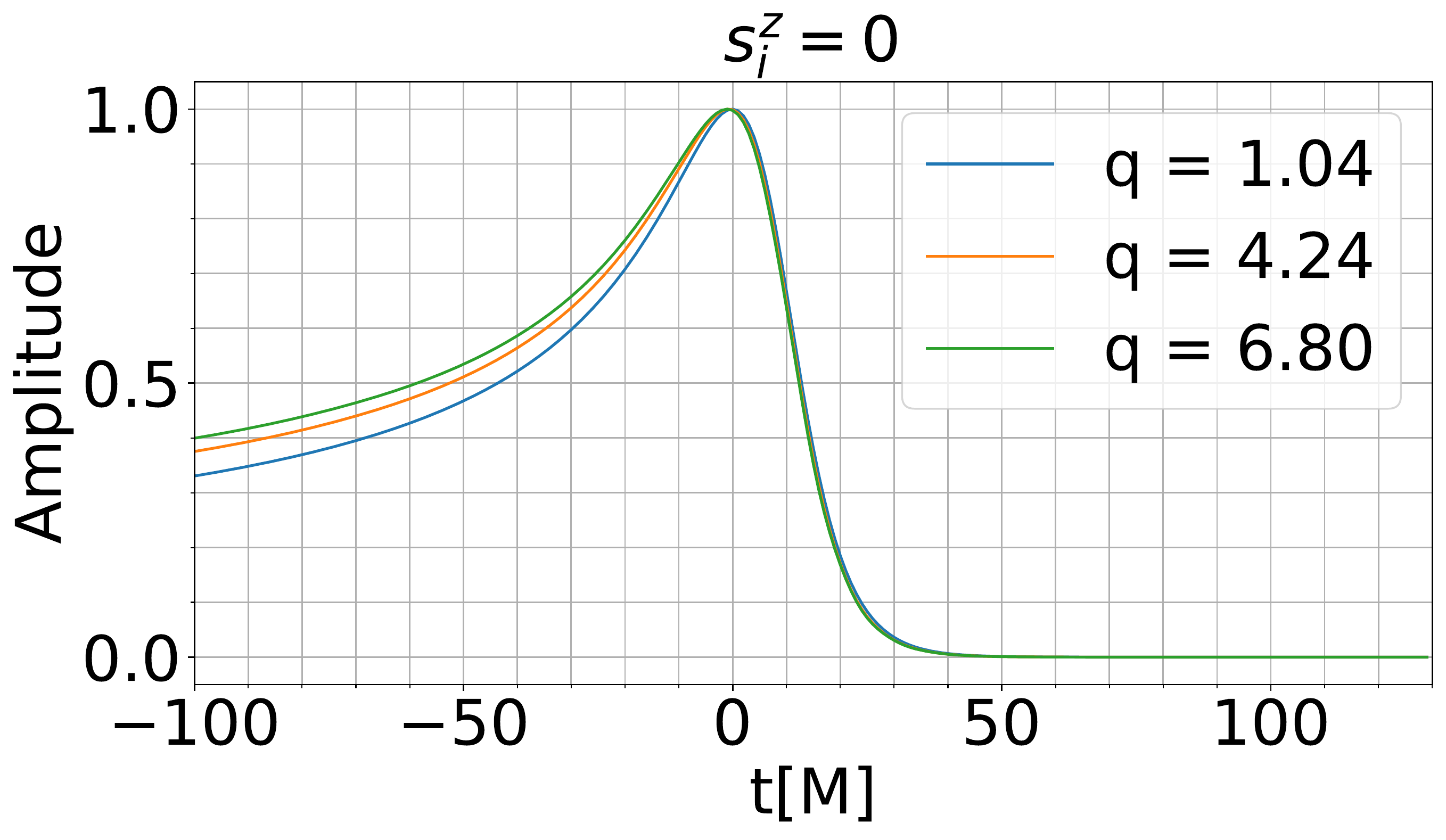}
}
\centerline{
\includegraphics[width=0.33\linewidth]{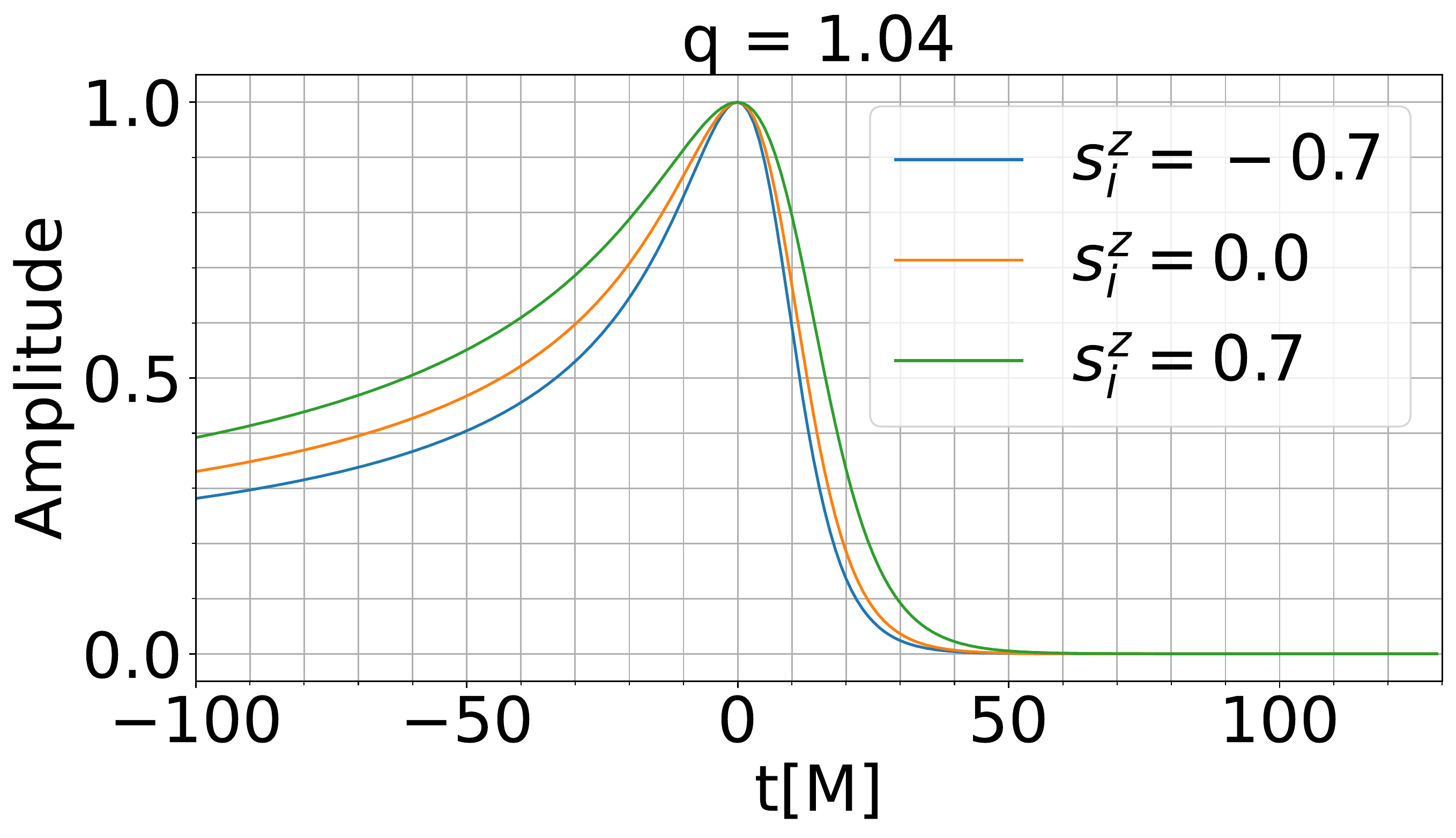}
\includegraphics[width=0.33\linewidth]{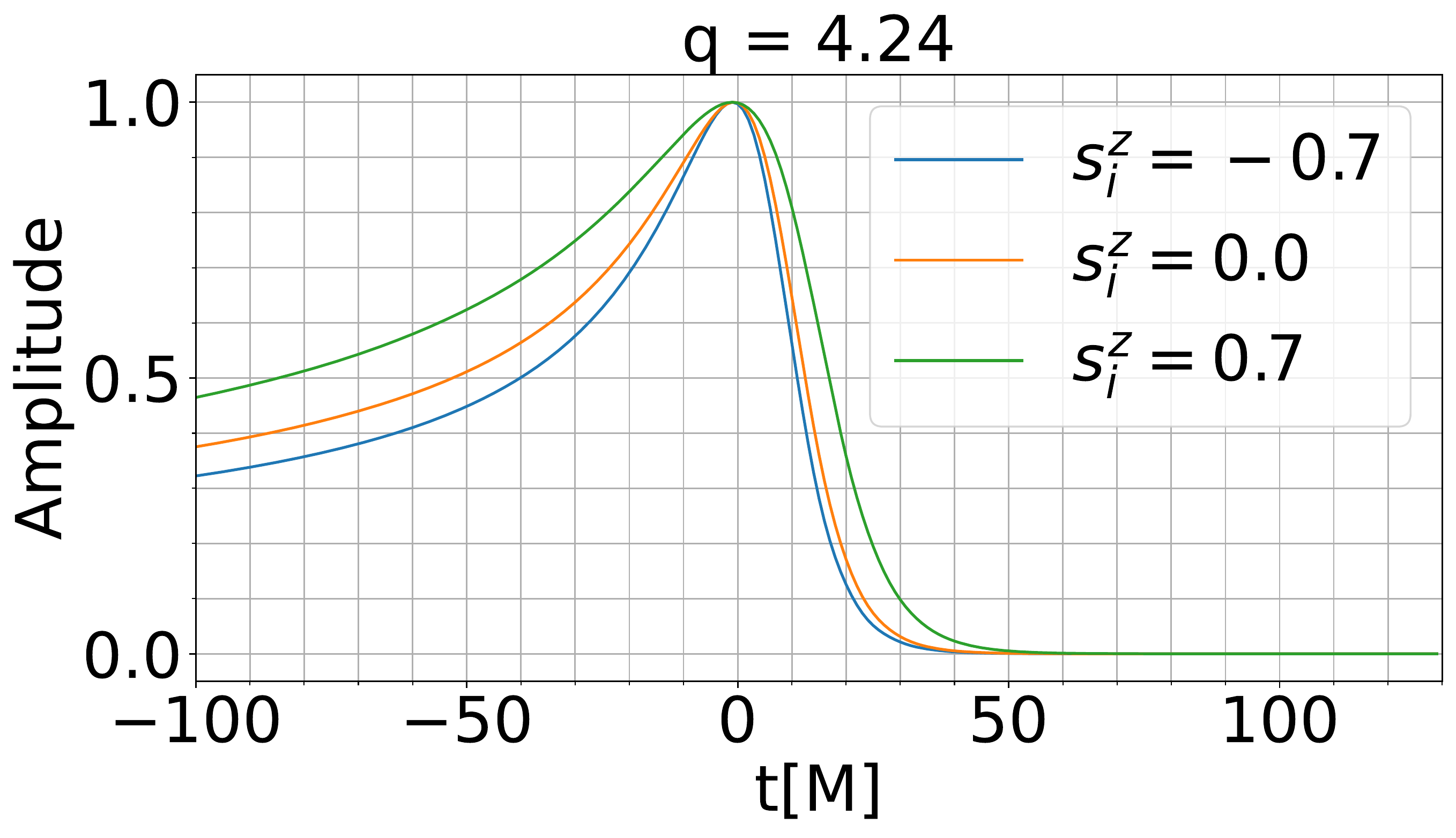}
\includegraphics[width=0.33\linewidth]{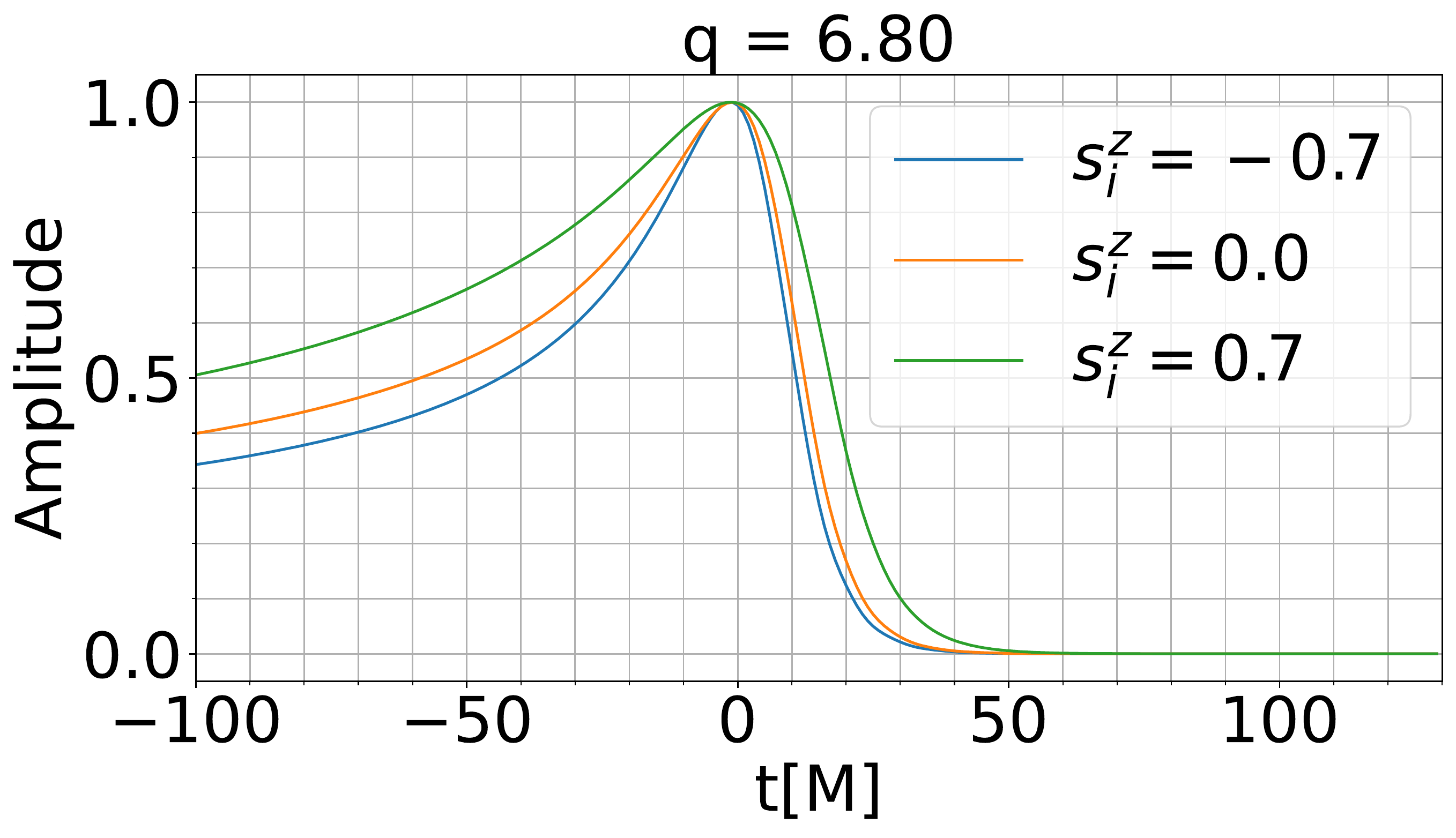}
} 
\caption{\textbf{Top panel} Near universal behaviour in 
the dynamical evolution of quasi-circular, non-spinning 
binary black hole mergers in the vicinity of 
merger \((t=0\textrm{M})\). \textbf{Bottom panels} Individual 
spins modulate the waveform amplitude, and drive binary 
black holes to merger in distinct ways, namely, rapid 
merger (left panel) and delayed merger (right panel). 
Notice the distinct features of the normalized amplitude 
for each system near merger. These subtle differences 
in waveform dynamics are highly non-trivial to capture 
by semi-analytical waveform models, though AI can accurately 
learn and predict these properties in a data-driven 
fashion.}
\label{fig:spin_dynamics}
\end{figure*}

To get a visual representation of the type of 
dynamics that our AI model needs to capture, we 
present in Figure~\ref{fig:spin_dynamics} the 
normalized waveform amplitude of the binary black 
hole systems considered in Figure~\ref{fig:examples}. 
Key points to extract from these results include: 

\begin{itemize}
    \item Top panel: Quasi-circular, non-spinning, binary 
    black holes display a well known universal behaviour 
    in the vicinity of merger. These physical properties 
    facilitate the training of AI models for these types of 
    systems. 
    \item Bottom panels: We notice the role individual 
    spins play in modulating 
    the waveform 
    amplitude, and driving the systems to merger. These 
    physical properties are one of the most 
    challenging features to capture for waveform 
    modeling experts who aim to accurately describe the late-time evolution of spinning, non-precessing 
    binary black hole mergers. In this study we have 
    demonstrated that AI may accomplish such a task in 
    data-driven manner.
\end{itemize}

We have quantified the accuracy of our model's predictions 
by computing the overlap, \({\cal{O}}  (h_t,\,h_p)\), 
between the target waveform $h_t$ and the 
predicted waveform $h_p$

\begin{eqnarray}
\label{over}
{\cal{O}}  (h_t,\,h_p)= \underset{ t_c\, \phi_c}{\mathrm{max}}\left(\hat{h}_t|\hat{h}_p{[t_c,\,  \phi_c]}\right)\,,
\nonumber\\ \quad{\rm with}\quad \hat{h}_t=h_t\,\left(h_t | h_t\right)^{-1/2}\,,
\end{eqnarray}

\noindent where \(\hat{h}_p{[t_c,\,  \phi_c]}\)  
indicates that the normalized waveform \(\hat{h}_p\) 
has been time- 
and phase-shifted. Hence, the overlap $\cal{O}$ lies in 
$[0,1]$, reaching the maximum value of $1$ for a perfect 
match. To visualize our findings, we first recast 
the parameter space \((q, s_1^z, s_2^z)\) into 
symmetric mass ratio $\eta$ and effective spin $\sigma_{\textrm{eff}}$ using the relations

\begin{equation}
   \label{recast}
    \eta = \frac{q}{(1+q)^2} \quad \mathrm{and}\quad
    \sigma_{\textrm{eff}} = \frac{q s^z_1 + s^z_2}{1 + q}\,.
\end{equation}

\begin{figure*}[ht]
\centerline{
\includegraphics[width=1.0\linewidth]{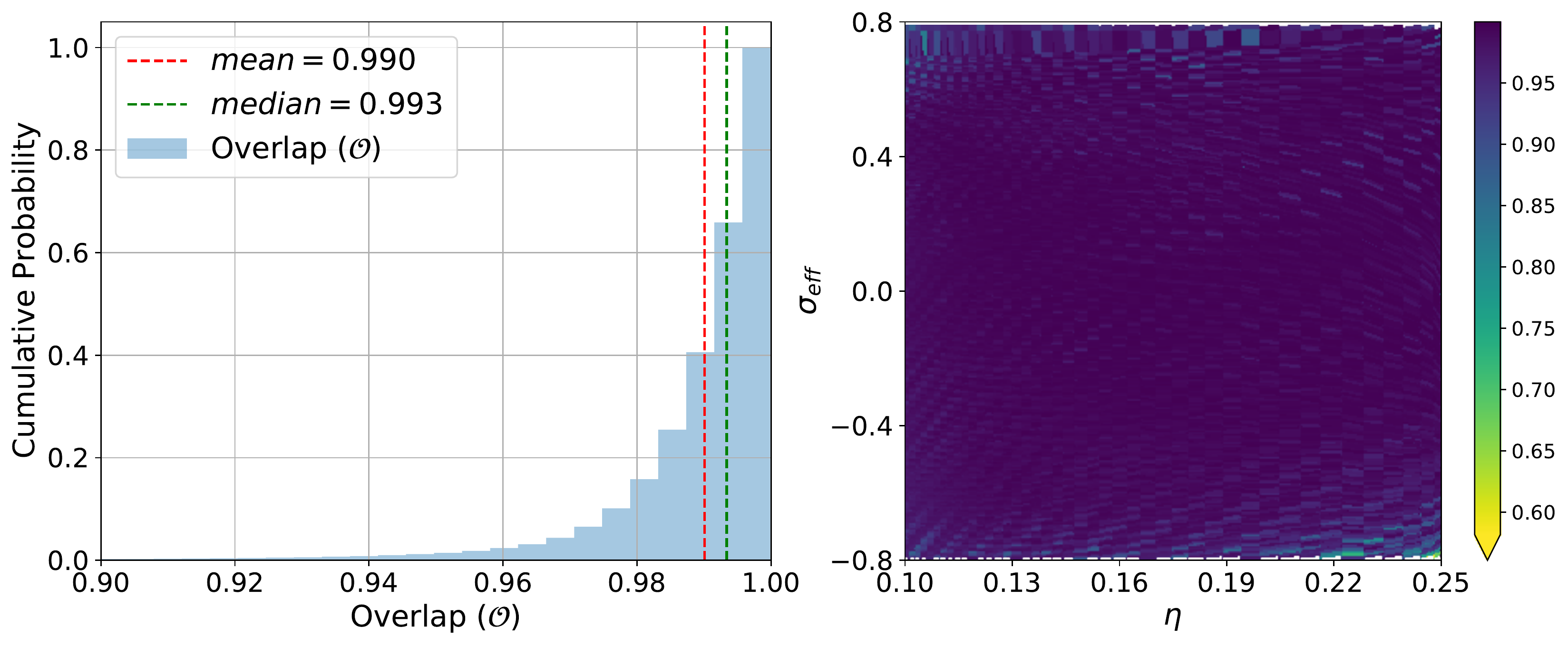}
} 
\centerline{
\includegraphics[width=1.0\linewidth]{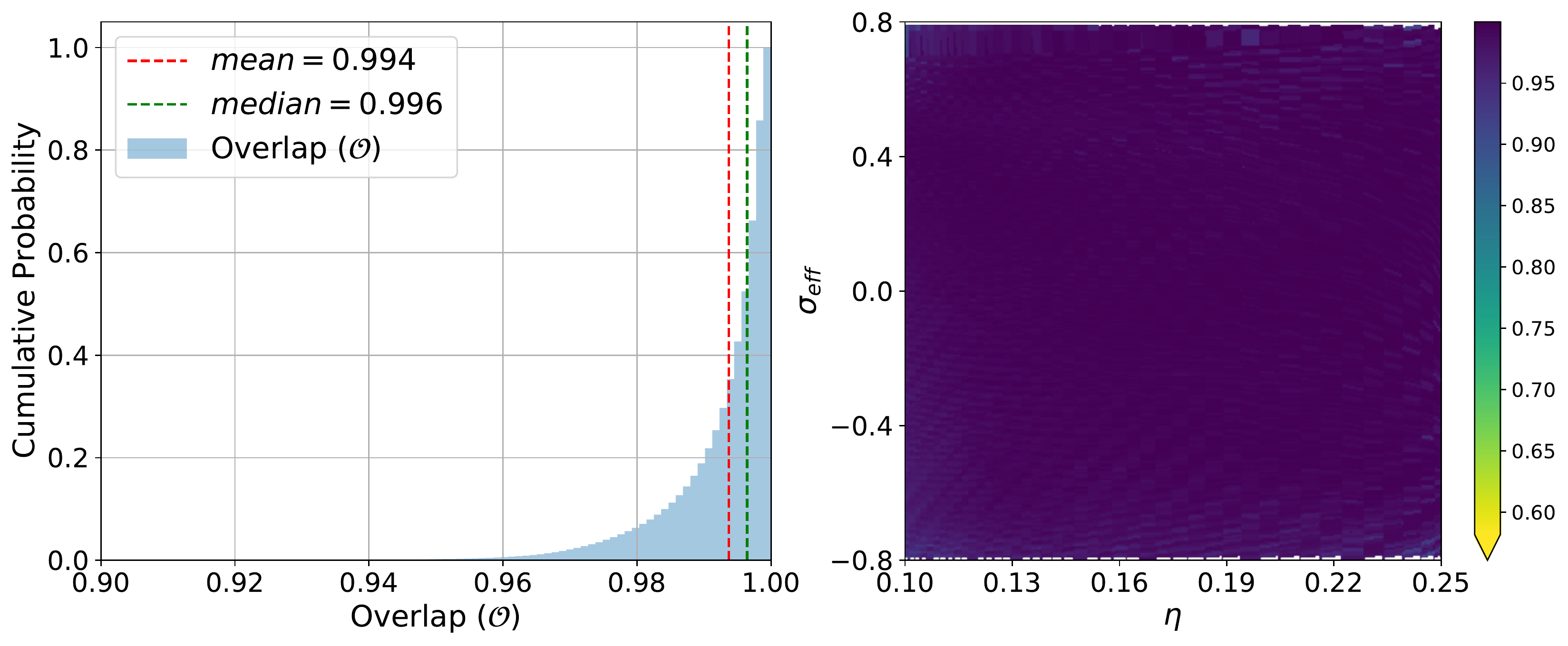}
} 
\centerline{
\includegraphics[width=1.0\linewidth]{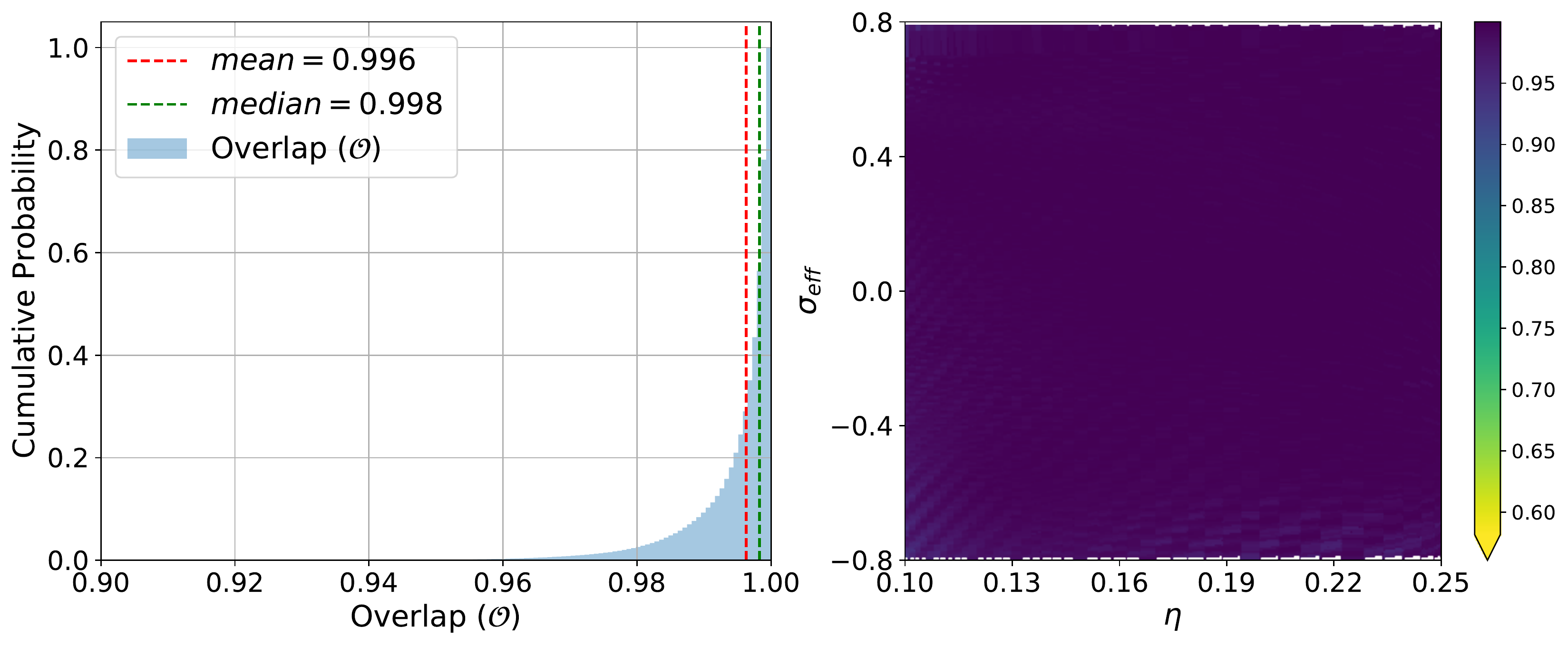}
} 
\caption{\textbf{Left column} Cumulative distribution 
of overlaps between target and predicted 
waveforms. From top to bottom, we present 
results for our AI model predicting the 
waveform evolution from \(t=\{-100\textrm{M}, -80\textrm{M}, -60\textrm{M}\}\), respectively. \textbf{Right column} 
Heatmap of the overlap distribution over the 
entire test set. We present results in terms of the 
symmetric mass-ratio and effective spin, \((\eta, \sigma_{\textrm{eff}})\), as defined in 
Equation~\eqref{recast}.}
\label{fig:histogram}
\end{figure*}

Using these conventions, we present 
overlap calculations between the target and 
predicted waveforms for the 
entire test data-set in Figure~\ref{fig:histogram}. 
To carry out these calculations, we 
used the plus and cross polarizations of the 
target waveforms spanning the range 
\(t\in[-5000\textrm{M}, 130\textrm{M}]\). 
Our target waveforms consist of input data spanning 
the range \(t\in[-5000\textrm{M}, -100\textrm{M}]\) 
and complemented with our predicted waveforms 
that span the range 
\(t\in[-100\textrm{M}, 130\textrm{M}]\). These 
calculations, presented in the 
top panels of Figure~\ref{fig:histogram}, 
indicate that 
both the mean and median overlaps 
${\cal{O}} > 0.99$,  and that less than 
$10\%$ of the test dataset has ${\cal{O}} < 0.98$. 
These outliers are localized at the edges of the 
parameter space, as shown in the top-right panel 
of Figure~\ref{fig:histogram}. In brief, 
our model predicts the late-inspiral, merger 
and ringdown waveform evolution in the time-range 
\([-100\textrm{M}, 130\textrm{M}]\). Since we 
sampled waveforms with 
a time step of \(2\textrm{M}\), this means that 
the model 
outputs 115 steps of waveform evolution.

We have also used our AI model to quantify the 
accuracy of its predictions from two additional 
initial times, namely \(t=\{-80\textrm{M}, -60\textrm{M}\}\). 
In these cases, the model outputs 
105 and 95 steps 
of waveform evolution, respectively. We present results for 
these cases in the mid and bottom panels of 
Figure~\ref{fig:histogram}. The overlap distributions 
for these cases are such that 

\begin{itemize}
    \item \(t=-80\textrm{M}\): median and 
    mean overlaps ${\cal{O}} > 0.994$, with 
    less than $6.1\%$ of the test dataset with ${\cal{O}} < 0.98$.
    \item \(t=-60\textrm{M}\): median and 
    mean overlaps ${\cal{O}} > 0.996$, with 
    less than $2.4\%$ of the test dataset with ${\cal{O}} < 0.98$.
\end{itemize}

\noindent We provide additional results that may be 
explored interactively in the website~\cite{asad_interactive}.
We see a progressive degradation in overlaps as 
we increase the target interval from \(t\in[-60\textrm{M}, 
130\textrm{M}]\) to \(t\in[-100\textrm{M}, 130\textrm{M}]\). To 
explore the cause of this effect further, we trained three more 
models tasked with predicting only the segments \([-80\textrm{M}, 
130\textrm{M}]\), \([-60\textrm{M}, 130\textrm{M}]\), and 
\([-50\textrm{M}, 130\textrm{M}]\) respectively.  Let's call these
models M80, M60, and M50 respectively. Then we noticed that the 
performance of M60 was slightly worse than M80 when predicting the
same segment \([-60\textrm{M}, 130\textrm{M}]\), and so on. 
This hints at some of the loss 
in performance coming from the margin effect, 
i.e., $-60\textrm{M}$ is at 
the margin during training for M60 but not for M80, etc. However, 
these small variations are hard to quantify due to inherent 
stochasticity of training deep neural networks. But more 
importantly, the most significant degradation in performance came 
from increasing the prediction span from \([-60\textrm{M}, 
130\textrm{M}]\) to \([-80\textrm{M}, 130\textrm{M}]\), and 
similarly from \([-80\textrm{M}, 130\textrm{M}]\) to 
\([-100\textrm{M}, 130\textrm{M}]\).

\begin{figure*}
\centerline{
\includegraphics[width=.5\linewidth]{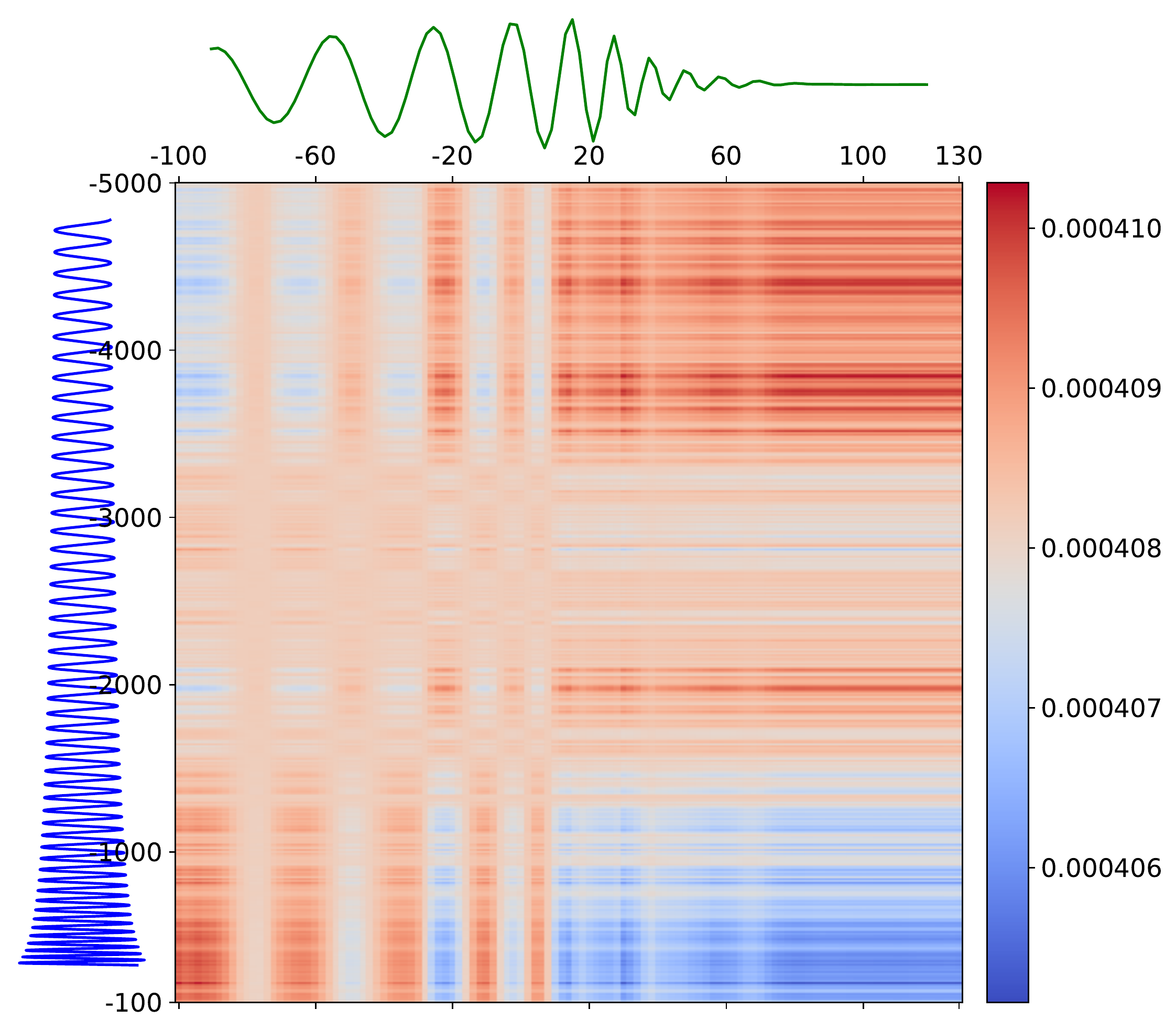}
\includegraphics[width=.47\linewidth]{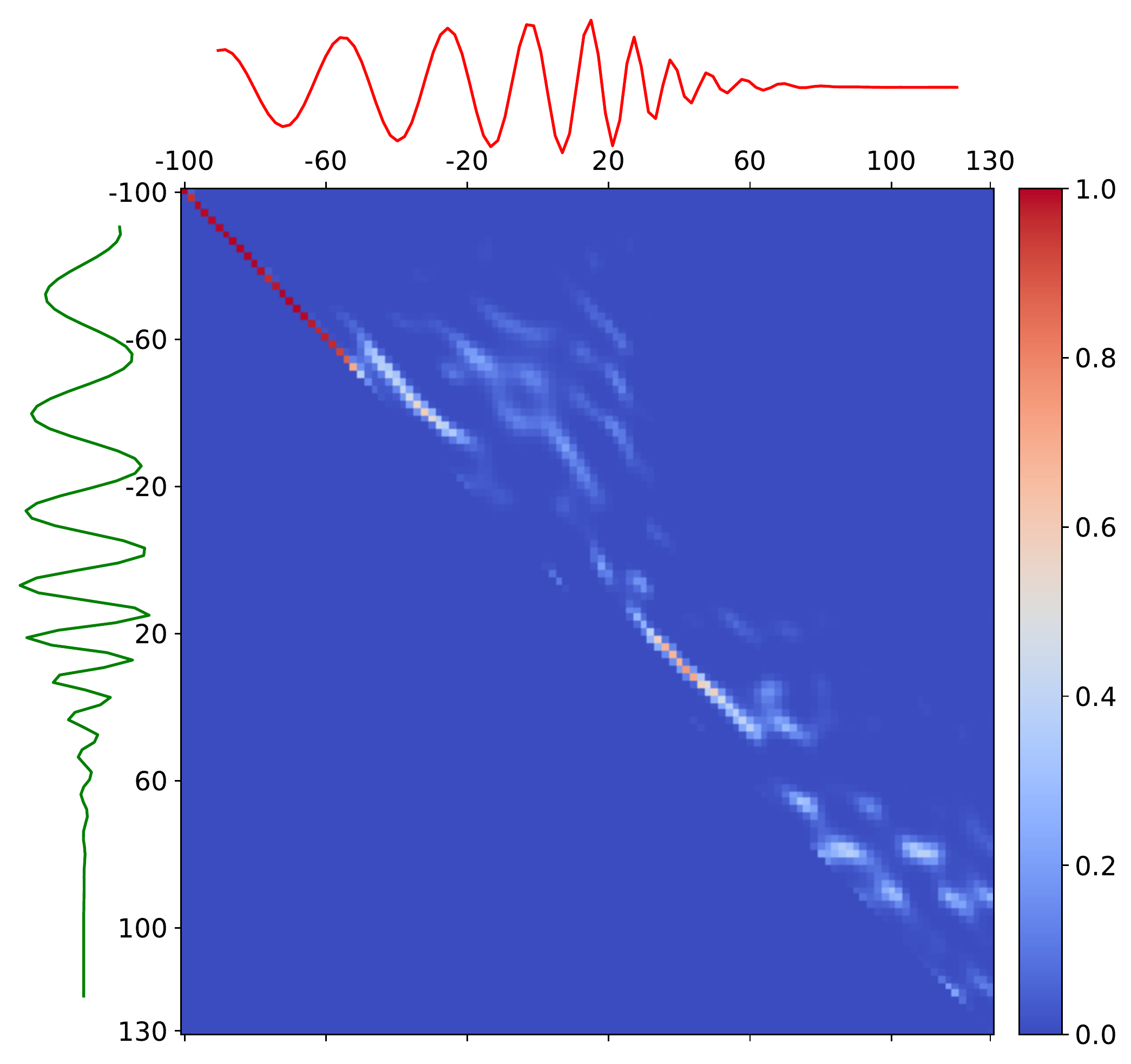}
} 
\caption{\textbf{Left panel} Heatmap for one of the 
twelve \textit{cross attention} heads showing which parts of the input waveform (shown in blue on the left) the decoder 
is paying attention to when predicting the output 
at any particular timestep (shown in green at the top). 
\textbf{Right panel} Heatmap showing one of the 
\textit{self attention} heads of the decoder.}
\label{fig:attention}
\end{figure*}

\noindent\textbf{Interpretability:} A nice side-effect 
and a major advantage of the attention mechanism is 
that it enables us to visualize and try to interpret 
what is happening inside the model. Looking 
at Equations~\eqref{eq1} and~\eqref{eq2} we notice 
that the coefficients $w_{ij}$ form a $t \times n$ 
matrix $A$. The $i^{th}$ row of $A$ consists of the 
attention scores over all the input vectors 
$\{x_1, x_2, x_3, ..., x_n\}$ when producing 
the $i^{th}$ output $h_i$, and hence each row sums 
up to $1$. Therefore visualizing the $i^{th}$ row 
of matrix $A$ shows which parts of the 
input $\{x_1, x_2, x_3, ..., x_n\}$ the model was 
``paying attention'' to when generating the 
$i^{th}$ output $h_i$. Visualizing the whole matrix 
$A$ then summarizes where the model was ``looking at" 
when generating each time-step of the output. 

In this vein, we visualize the self-attention 
and cross-attention score matrices of the decoder module 
when generating the predictions for a sample waveform 
with parameters \(\{q, s^z_1, s^z_2\}=\{6.8, 0.718, 0.574\}\) 
in Figure~\ref{fig:attention}. Therein we present 
results for one of the 12 attention heads from our 
model's decoder. We present additional results for the 
other attention heads in Appendix~\ref{sec:ap2}.

The left panel of Figure~\ref{fig:attention} 
shows the transpose of the cross-attention score matrix. 
Each column $j$ shows which parts of the input 
waveform segment 
(\(t\in[-5000\,\textrm{M}, -100\,\textrm{M}]\)) the model 
was paying attention to when predicting the 
$j^{th}$ time-step of the target waveform 
segment (\(t_j\in[-100\,\textrm{M}, 130\,\textrm{M}]\)). 
For reference, we also plot the input waveform segment 
and the predicted waveform segment to the left and 
top of the matrix, respectively. We see that for 
the late-inspiral and merger phases of the prediction, 
the model is paying a diffused form of attention to 
the whole input segment, occasionally flip-flopping, 
i.e., paying more attention to the late-inspiral rather 
than early inspiral and vice versa. 
However, when predicting the ringdown, all of the 
attention gets focused towards the early 
inspiral of the input segment.

The right panel of Figure~\ref{fig:attention} shows 
the transpose of the self-attention matrix. Since 
predictions are generated autoregressively, the 
self-attention here is causal, i.e., at any given 
time-step $t=j$, the model cannot pay attention to 
future time-steps $t > j$. Consequently this matrix 
is upper triangular with a strong correlation between 
adjacent time-steps, thus mostly diagonal.

The results in Figure~\ref{fig:attention} provide 
a glimpse of the activity happening within our trained
AI model that is 
responsible for accurate and reliable forecasting 
predictions. For the interested readers, we provide 
in the website~\cite{asad_interactive} additional interactive results 
to enhance our intuition into how our AI model behaves 
for different astrophysical configurations.

\section{Conclusions}
\label{sec:end}

\noindent We have designed an AI model that is 
capable of learning 
and predicting the late-inspiral, merger and 
ringdown evolution of quasi-circular, spinning, 
non-precessing binary black hole mergers. The data-driven 
methodology used to create these AI tools demonstrates that 
AI can learn and accurately describe the 
plus and cross polarizations of numerical 
relativity waveforms when we feed input signals that 
contain information up to \(-100M\) before the 
merger event (defined as the amplitude peak of the waveform 
signal). We have also demonstrated that our AI model 
may forecast the waveform evolution starting at some 
other initial time $t_i$. In this study we presented 
quantitative results for the cases \(t_i=-80\textrm{M}\) and 
\(t_i=-60\textrm{M}\). In all these cases, the mean 
and median overlap between target and predicted 
waveforms is \({\cal{O}}\geq0.99\).

We have also explored visualizing several components in 
our AI model (i.e., the various attentions heads) that 
are responsible for data-driven decision making 
and waveform forecasting. In particular, we generated 
visualizations to see which components of the input are 
responsible for the prediction of the pre-merger, merger 
and ringdown pieces of our predicted waveforms. We have 
made available an interactive website where users can 
explore these results in further detail for a variety of
astrophysical systems. We expect that this approach 
persuades other researchers 
to go a step beyond and try to understand how AI models 
make predictions, and will help advance other efforts 
on creating interpretable AI models.

\begin{acknowledgments}
\label{ack}
\noindent A.K. and E.A.H. gratefully acknowledge National 
Science Foundation (NSF) awards OAC-1931561 and 
OAC-1934757. This research used resources of the Argonne 
Leadership Computing Facility, which is a DOE Office of 
Science User Facility supported under Contract DE-AC02-06CH11357.

E.A.H. gratefully acknowledges the Innovative and 
Novel Computational Impact on Theory and Experiment 
project `Multi-Messenger Astrophysics 
at Extreme Scale in Summit'. This research used 
resources of the Oak Ridge Leadership Computing Facility, 
which is a DOE Office of Science User Facility 
supported under contract no. DE-AC05-00OR22725. 
This work utilized resources supported by the 
NSF's Major Research Instrumentation program, 
the HAL cluster (grant no. OAC-1725729), 
as well as the University of Illinois at 
Urbana-Champaign. We thank \texttt{NVIDIA} for 
their continued support. 
\end{acknowledgments}

\bibliography{book_references}

\begin{thebibliography}{10}

\bibitem{geodf:2017a}
D.~George and E.~A. Huerta, ``Deep neural networks to enable real-time
  multimessenger astrophysics,'' {\em Phys. Rev. D}, vol.~97, p.~044039, Feb
  2018.

\bibitem{GEORGE201864}
D.~George and E.~Huerta, ``Deep learning for real-time gravitational wave
  detection and parameter estimation: Results with advanced ligo data,'' {\em
  Physics Letters B}, vol.~778, pp.~64 -- 70, 2018.

\bibitem{2018GN}
H.~{Gabbard}, M.~{Williams}, F.~{Hayes}, and C.~{Messenger}, ``{Matching
  Matched Filtering with Deep Networks for Gravitational-Wave Astronomy},''
  {\em Physical Review Letters}, vol.~120, p.~141103, Apr. 2018.

\bibitem{2020arXiv200914611S}
V.~{Skliris}, M.~R.~K. {Norman}, and P.~J. {Sutton}, ``{Real-Time Detection of
  Unmodeled Gravitational-Wave Transients Using Convolutional Neural
  Networks},'' {\em arXiv e-prints}, p.~arXiv:2009.14611, Sept. 2020.

\bibitem{Lin:2020aps}
Y.-C. Lin and J.-H.~P. Wu, ``Detection of gravitational waves using bayesian
  neural networks,'' {\em Phys. Rev. D}, vol.~103, p.~063034, Mar 2021.

\bibitem{Wang:2019zaj}
H.~Wang, S.~Wu, Z.~Cao, X.~Liu, and J.-Y. Zhu, ``{Gravitational-wave signal
  recognition of LIGO data by deep learning},'' {\em Phys. Rev. D}, vol.~101,
  no.~10, p.~104003, 2020.

\bibitem{Fan:2018vgw}
X.~Fan, J.~Li, X.~Li, Y.~Zhong, and J.~Cao, ``{Applying deep neural networks to
  the detection and space parameter estimation of compact binary coalescence
  with a network of gravitational wave detectors},'' {\em Sci. China Phys.
  Mech. Astron.}, vol.~62, no.~6, p.~969512, 2019.

\bibitem{Li:2017chi}
X.-R. Li, G.~Babu, W.-L. Yu, and X.-L. Fan, ``{Some optimizations on detecting
  gravitational wave using convolutional neural network},'' {\em Front. Phys.
  (Beijing)}, vol.~15, no.~5, p.~54501, 2020.

\bibitem{Deighan:2020gtp}
D.~S. {Deighan}, S.~E. {Field}, C.~D. {Capano}, and G.~{Khanna},
  ``{Genetic-algorithm-optimized neural networks for gravitational wave
  classification},'' {\em arXiv e-prints}, p.~arXiv:2010.04340, Oct. 2020.

\bibitem{Miller:2019jtp}
A.~L. Miller {\em et~al.}, ``{How effective is machine learning to detect long
  transient gravitational waves from neutron stars in a real search?},'' {\em
  Phys. Rev. D}, vol.~100, no.~6, p.~062005, 2019.

\bibitem{Krastev:2019koe}
P.~G. Krastev, ``{Real-Time Detection of Gravitational Waves from Binary
  Neutron Stars using Artificial Neural Networks},'' {\em Phys. Lett. B},
  vol.~803, p.~135330, 2020.

\bibitem{2020PhRvD.102f3015S}
M.~B. {Sch{\"a}fer}, F.~{Ohme}, and A.~H. {Nitz}, ``{Detection of
  gravitational-wave signals from binary neutron star mergers using machine
  learning},'' {\em \prd}, vol.~102, p.~063015, Sept. 2020.

\bibitem{Dreissigacker:2020xfr}
C.~Dreissigacker and R.~Prix, ``{Deep-Learning Continuous Gravitational Waves:
  Multiple detectors and realistic noise},'' {\em Phys. Rev. D}, vol.~102,
  no.~2, p.~022005, 2020.

\bibitem{Adam:2018prd}
A.~{Rebei}, E.~A. {Huerta}, S.~{Wang}, S.~{Habib}, R.~{Haas}, D.~{Johnson}, and
  D.~{George}, ``{Fusing numerical relativity and deep learning to detect
  higher-order multipole waveforms from eccentric binary black hole mergers},''
  {\em \prd}, vol.~100, p.~044025, Aug 2019.

\bibitem{Dreissigacker:2019edy}
C.~Dreissigacker, R.~Sharma, C.~Messenger, R.~Zhao, and R.~Prix,
  ``{Deep-Learning Continuous Gravitational Waves},'' {\em Phys. Rev. D},
  vol.~100, no.~4, p.~044009, 2019.

\bibitem{2020PhRvD.101f4009B}
B.~{Beheshtipour} and M.~A. {Papa}, ``{Deep learning for clustering of
  continuous gravitational wave candidates},'' {\em \prd}, vol.~101, p.~064009,
  Mar. 2020.

\bibitem{2021arXiv210810715S}
M.~B. {Sch{\"a}fer} and A.~H. {Nitz}, ``{From One to Many: A Deep Learning
  Coincident Gravitational-Wave Search},'' {\em arXiv e-prints},
  p.~arXiv:2108.10715, Aug. 2021.

\bibitem{shen2019denoising}
H.~Shen, D.~George, E.~A. Huerta, and Z.~Zhao, ``Denoising gravitational waves
  with enhanced deep recurrent denoising auto-encoders,'' in {\em ICASSP
  2019-2019 IEEE International Conference on Acoustics, Speech and Signal
  Processing (ICASSP)}, pp.~3237--3241, IEEE, 2019.

\bibitem{Wei:2019zlc}
W.~Wei and E.~A. Huerta, ``{Gravitational Wave Denoising of Binary Black Hole
  Mergers with Deep Learning},'' {\em Phys. Lett.}, vol.~B800, p.~135081, 2020.

\bibitem{PhysRevResearch.2.033066}
R.~Ormiston, T.~Nguyen, M.~Coughlin, R.~X. Adhikari, and E.~Katsavounidis,
  ``Noise reduction in gravitational-wave data via deep learning,'' {\em Phys.
  Rev. Research}, vol.~2, p.~033066, Jul 2020.

\bibitem{Shen:2019vep}
H.~Shen, E.~Huerta, E.~O’Shea, P.~Kumar, and Z.~Zhao,
  ``Statistically-informed deep learning for gravitational wave parameter
  estimation,'' {\em Machine Learning: Science and Technology}, vol.~3, no.~1,
  p.~015007, 2021.

\bibitem{Gabbard:2019rde}
H.~Gabbard, C.~Messenger, I.~S. Heng, F.~Tonolini, and R.~Murray-Smith,
  ``{Bayesian parameter estimation using conditional variational autoencoders
  for gravitational-wave astronomy},'' {\em Nature Physics}, 2021.

\bibitem{Chua:2019wwt}
A.~J. Chua and M.~Vallisneri, ``{Learning Bayesian posteriors with neural
  networks for gravitational-wave inference},'' {\em Phys. Rev. Lett.},
  vol.~124, no.~4, p.~041102, 2020.

\bibitem{Green:2020hst}
S.~R. Green, C.~Simpson, and J.~Gair, ``Gravitational-wave parameter estimation
  with autoregressive neural network flows,'' {\em Phys. Rev. D}, vol.~102,
  p.~104057, Nov 2020.

\bibitem{Green:2020dnx}
S.~R. Green and J.~Gair, ``Complete parameter inference for gw150914 using deep
  learning,'' {\em Machine Learning: Science and Technology}, vol.~2, no.~3,
  p.~03LT01, 2021.

\bibitem{2021arXiv210612594D}
M.~Dax, S.~R. Green, J.~Gair, J.~H. Macke, A.~Buonanno, and B.~Sch{\"o}lkopf,
  ``Real-time gravitational wave science with neural posterior estimation,''
  {\em Physical review letters}, vol.~127, no.~24, p.~241103, 2021.

\bibitem{Khan:2020fso}
S.~Khan and R.~Green, ``{Gravitational-wave surrogate models powered by
  artificial neural networks},'' {\em Phys. Rev. D}, vol.~103, no.~6,
  p.~064015, 2021.

\bibitem{PhysRevLett.122.211101}
A.~J.~K. Chua, C.~R. Galley, and M.~Vallisneri, ``Reduced-order modeling with
  artificial neurons for gravitational-wave inference,'' {\em Phys. Rev.
  Lett.}, vol.~122, p.~211101, May 2019.

\bibitem{2020arXiv201203963W}
W.~{Wei}, E.~A. {Huerta}, M.~{Yun}, N.~{Loutrel}, M.~A. {Shaikh}, P.~{Kumar},
  R.~{Haas}, and V.~{Kindratenko}, ``{Deep Learning with Quantized Neural
  Networks for Gravitational-wave Forecasting of Eccentric Compact Binary
  Coalescence},'' {\em \apj}, vol.~919, p.~82, Oct. 2021.

\bibitem{Wei:2020sfz}
W.~Wei and E.~A. Huerta, ``{Deep learning for gravitational wave forecasting of
  neutron star mergers},'' {\em Phys. Lett. B}, vol.~816, p.~136185, 2021.

\bibitem{2021PhRvD.104f2004Y}
H.~{Yu}, R.~X. {Adhikari}, R.~{Magee}, S.~{Sachdev}, and Y.~{Chen}, ``{Early
  warning of coalescing neutron-star and neutron-star-black-hole binaries from
  the nonstationary noise background using neural networks},'' {\em \prd},
  vol.~104, p.~062004, Sept. 2021.

\bibitem{KHAN2020135628}
A.~Khan, E.~Huerta, and A.~Das, ``Physics-inspired deep learning to
  characterize the signal manifold of quasi-circular, spinning, non-precessing
  binary black hole mergers,'' {\em Physics Letters B}, vol.~808, p.~135628,
  2020.

\bibitem{2021PhLB..81236029W}
W.~{Wei}, A.~{Khan}, E.~A. {Huerta}, X.~{Huang}, and M.~{Tian}, ``{Deep
  learning ensemble for real-time gravitational wave detection of spinning
  binary black hole mergers},'' {\em Physics Letters B}, vol.~812, p.~136029,
  Jan. 2021.

\bibitem{2020arXiv201208545H}
E.~Huerta, A.~Khan, X.~Huang, M.~Tian, M.~Levental, R.~Chard, W.~Wei,
  M.~Heflin, D.~S. Katz, V.~Kindratenko, {\em et~al.}, ``Accelerated, scalable
  and reproducible ai-driven gravitational wave detection,'' {\em Nature
  Astronomy}, vol.~5, no.~10, pp.~1062--1068, 2021.

\bibitem{Nat_Rev_2019_Huerta}
E.~A. {Huerta}, G.~{Allen}, I.~{Andreoni}, J.~M. {Antelis}, E.~{Bachelet},
  G.~B. {Berriman}, F.~B. {Bianco}, R.~{Biswas}, M.~{Carrasco Kind},
  K.~{Chard}, M.~{Cho}, P.~S. {Cowperthwaite}, Z.~B. {Etienne}, M.~{Fishbach},
  F.~{Forster}, D.~{George}, T.~{Gibbs}, M.~{Graham}, W.~{Gropp}, R.~{Gruendl},
  A.~{Gupta}, R.~{Haas}, S.~{Habib}, E.~{Jennings}, M.~W.~G. {Johnson},
  E.~{Katsavounidis}, D.~S. {Katz}, A.~{Khan}, V.~{Kindratenko}, W.~T.~C.
  {Kramer}, X.~{Liu}, A.~{Mahabal}, Z.~{Marka}, K.~{McHenry}, J.~M. {Miller},
  C.~{Moreno}, M.~S. {Neubauer}, S.~{Oberlin}, A.~R. {Olivas}, D.~{Petravick},
  A.~{Rebei}, S.~{Rosofsky}, M.~{Ruiz}, A.~{Saxton}, B.~F. {Schutz},
  A.~{Schwing}, E.~{Seidel}, S.~L. {Shapiro}, H.~{Shen}, Y.~{Shen}, L.~P.
  {Singer}, B.~M. {Sipocz}, L.~{Sun}, J.~{Towns}, A.~{Tsokaros}, W.~{Wei},
  J.~{Wells}, T.~J. {Williams}, J.~{Xiong}, and Z.~{Zhao}, ``{Enabling
  real-time multi-messenger astrophysics discoveries with deep learning},''
  {\em Nature Reviews Physics}, vol.~1, pp.~600--608, Oct. 2019.

\bibitem{huerta_book}
E.~A. Huerta and Z.~Zhao, {\em Advances in Machine and Deep Learning for
  Modeling and Real-Time Detection of Multi-messenger Sources}, pp.~1--27.
\newblock Singapore: Springer Singapore, 2020.

\bibitem{2020arXiv200503745C}
E.~Cuoco, J.~Powell, M.~Cavagli{\`a}, K.~Ackley, M.~Bejger, C.~Chatterjee,
  M.~Coughlin, S.~Coughlin, P.~Easter, R.~Essick, {\em et~al.}, ``Enhancing
  gravitational-wave science with machine learning,'' {\em Machine Learning:
  Science and Technology}, vol.~2, no.~1, p.~011002, 2020.

\bibitem{2020PhRvD.101h4024R}
S.~G. {Rosofsky} and E.~A. {Huerta}, ``{Artificial neural network subgrid
  models of 2D compressible magnetohydrodynamic turbulence},'' {\em \prd},
  vol.~101, p.~084024, Apr. 2020.

\bibitem{2019JCoPh.394...56Z}
Y.~{Zhu}, N.~{Zabaras}, P.-S. {Koutsourelakis}, and P.~{Perdikaris},
  ``{Physics-constrained deep learning for high-dimensional surrogate modeling
  and uncertainty quantification without labeled data},'' {\em Journal of
  Computational Physics}, vol.~394, pp.~56--81, Oct. 2019.

\bibitem{Anirudh9741}
R.~Anirudh, J.~J. Thiagarajan, P.-T. Bremer, and B.~K. Spears, ``Improved
  surrogates in inertial confinement fusion with manifold and cycle
  consistencies,'' {\em Proceedings of the National Academy of Sciences},
  vol.~117, no.~18, pp.~9741--9746, 2020.

\bibitem{dl_rg_lee}
J.~{Lee}, S.~H. {Oh}, K.~{Kim}, G.~{Cho}, J.~J. {Oh}, E.~J. {Son}, and H.~M.
  {Lee}, ``{Deep learning model on gravitational waveforms in merging and
  ringdown phases of binary black hole coalescences},'' {\em \prd}, vol.~103,
  p.~123023, June 2021.

\bibitem{PhysRevD.99.064045}
V.~Varma, S.~E. Field, M.~A. Scheel, J.~Blackman, L.~E. Kidder, and H.~P.
  Pfeiffer, ``Surrogate model of hybridized numerical relativity binary black
  hole waveforms,'' {\em Phys. Rev. D}, vol.~99, p.~064045, Mar 2019.

\bibitem{2017arXiv170603762V}
A.~{Vaswani}, N.~{Shazeer}, N.~{Parmar}, J.~{Uszkoreit}, L.~{Jones}, A.~N.
  {Gomez}, L.~{Kaiser}, and I.~{Polosukhin}, ``{Attention Is All You Need},''
  {\em arXiv e-prints}, p.~arXiv:1706.03762, June 2017.

\bibitem{asad_interactive}
A.~{Khan}, E.~A. {Huerta}, and H.~{Zheng}, ``{Interpretable AI forecasting for
  numerical relativity waveforms of quasi-circular, spinning, non-precessing
  binary black hole mergers}.''
  \url{https://khanx169.github.io/gw_forecasting/interactive_results.html},
  2021.
\newblock [Online from October 2021].

\end{thebibliography}
\bibliographystyle{ieeetr}
\clearpage 
\appendix

\section{Teacher Forcing}
\label{sec:ap1}

Our model is designed to 
predict the waveform evolution in the time range 
\(-100\textrm{M} \leq t \leq 130\textrm{M}\). During both training
and inference, we compute the loss and quantify the performance of the
model by comparing the entire predicted and ground truth waveforms in the
time segment \(-100\textrm{M} \leq t 
\leq 130\textrm{M}\).

\begin{figure*}[h!]
\centerline{
\includegraphics[width=.8\linewidth]{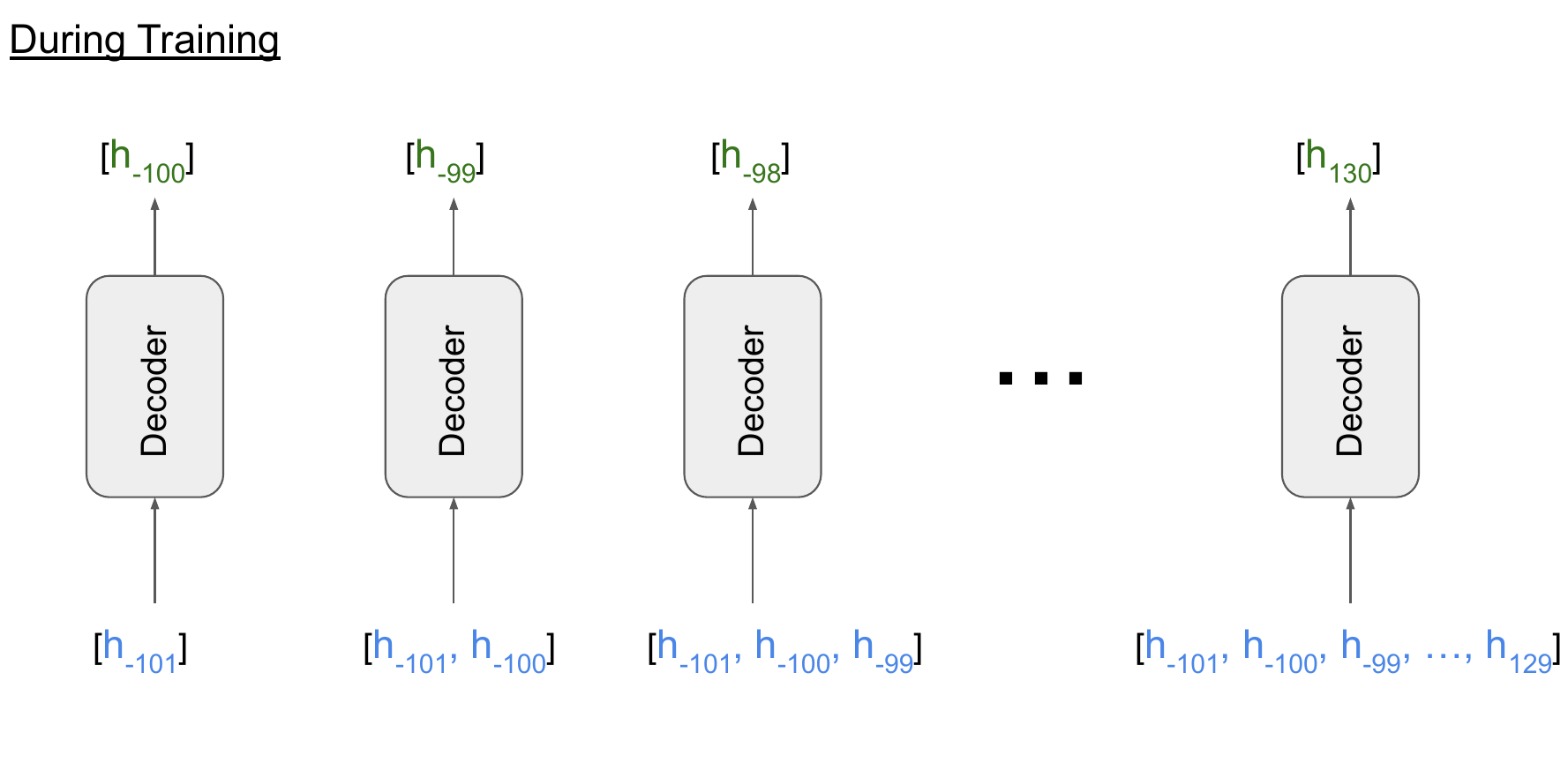}
}
\centerline{
\includegraphics[width=.8\linewidth]{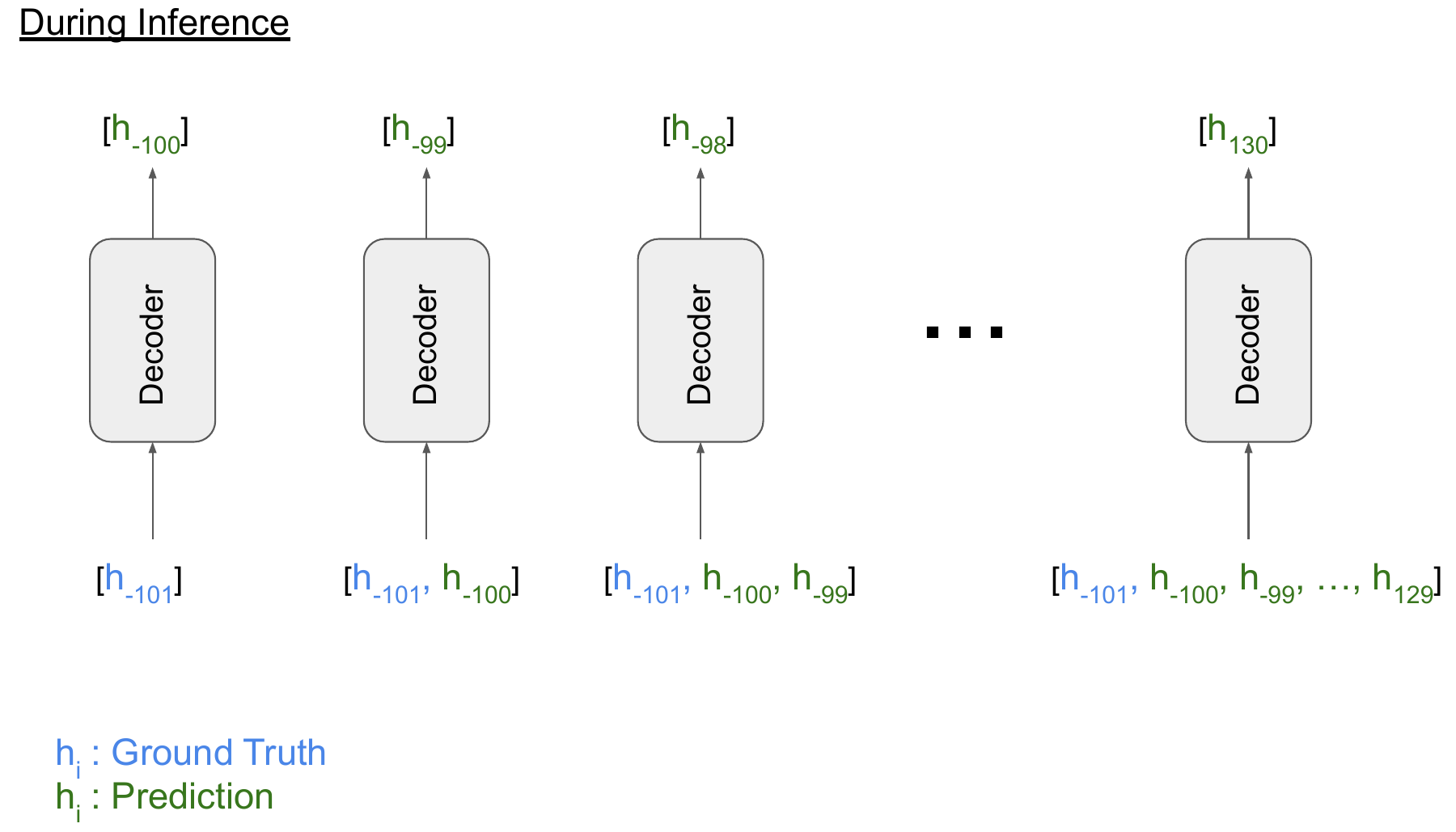}
}
\caption{A visual representation of \texttt{Teacher Forcing} approach \textbf{Top panel} Teacher Forcing is 
used during training; at each time step the decoder 
is fed the ground truth target values from the previous
time steps. \textbf{Bottom panel} During inference, Teacher Forcing is 
turned off, and at each time step the decoder is fed its own predicted values from the previous time steps.}
\label{fig:teacher_forcing}
\end{figure*}
 
During both 
training and inference we feed the input waveform 
covering the time-span \([-5000\textrm{M}, 
-100\textrm{M}]\) into the encoder. 
Additionally, during training we also employ \texttt{Teacher Forcing}, 
whereby at each time step the decoder 
is fed the ground truth target values from the previous
time steps, as illustrated 
in the top panel of Figure~\ref{fig:teacher_forcing}. This
methodology results in a more stable training and helps the model
converge faster. Finally, during inference we turn off \texttt{Teacher Forcing}
and instead feed the decoder its own predictions from the previous time-steps,
as illustrated 
in the bottom panel of Figure~\ref{fig:teacher_forcing}.

\section{Interpretability}
\label{sec:ap2}

We provide additional results for the 
12 attention heads that our AI model utilizes for 
the forecasting of numerical relativity waveforms. 
As in Figure~\ref{fig:attention}, 
we have produced these results for a binary black hole 
system with parameters 
\(\{q, s^z_1, s^z_2\}=\{6.8, 0.718, 0.574\}\). For 
additional results, we refer readers to the 
interactive website~\cite{asad_interactive}.

\begin{figure*}[h!]
\centerline{
\includegraphics[width=\linewidth]{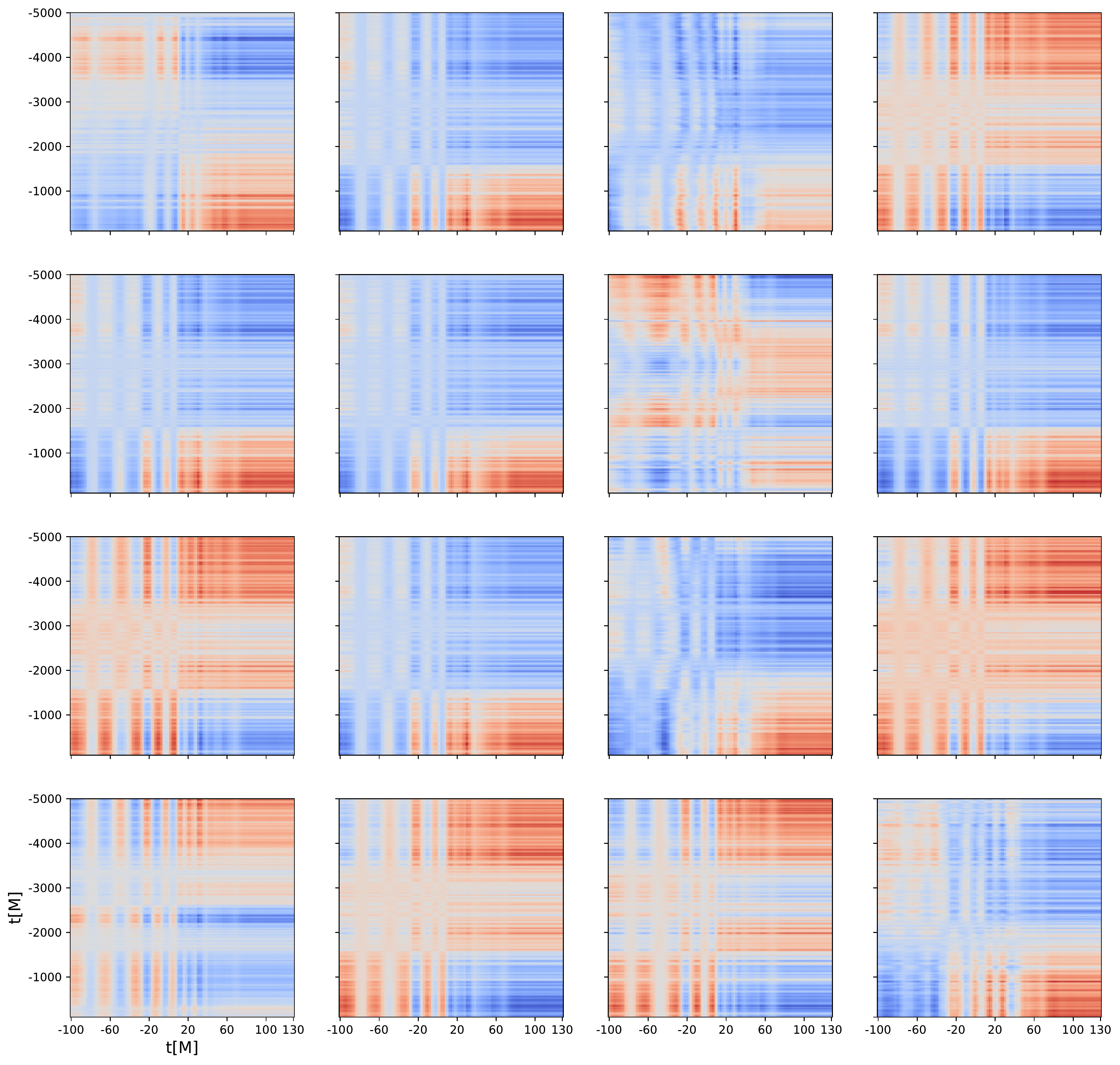}
} 
\caption{Response of all cross attention heads to 
a given input signal, indicating which parts of the input
waveform signal are taken into account to forecast the 
late-inspiral, merger and ringdown evolution. This 
behaviour is very consistent across the parameter 
space under consideration.}
\label{fig:all_cross_attention_heads}
\end{figure*}

\begin{figure*}[h!]
\centerline{
\includegraphics[width=\linewidth]{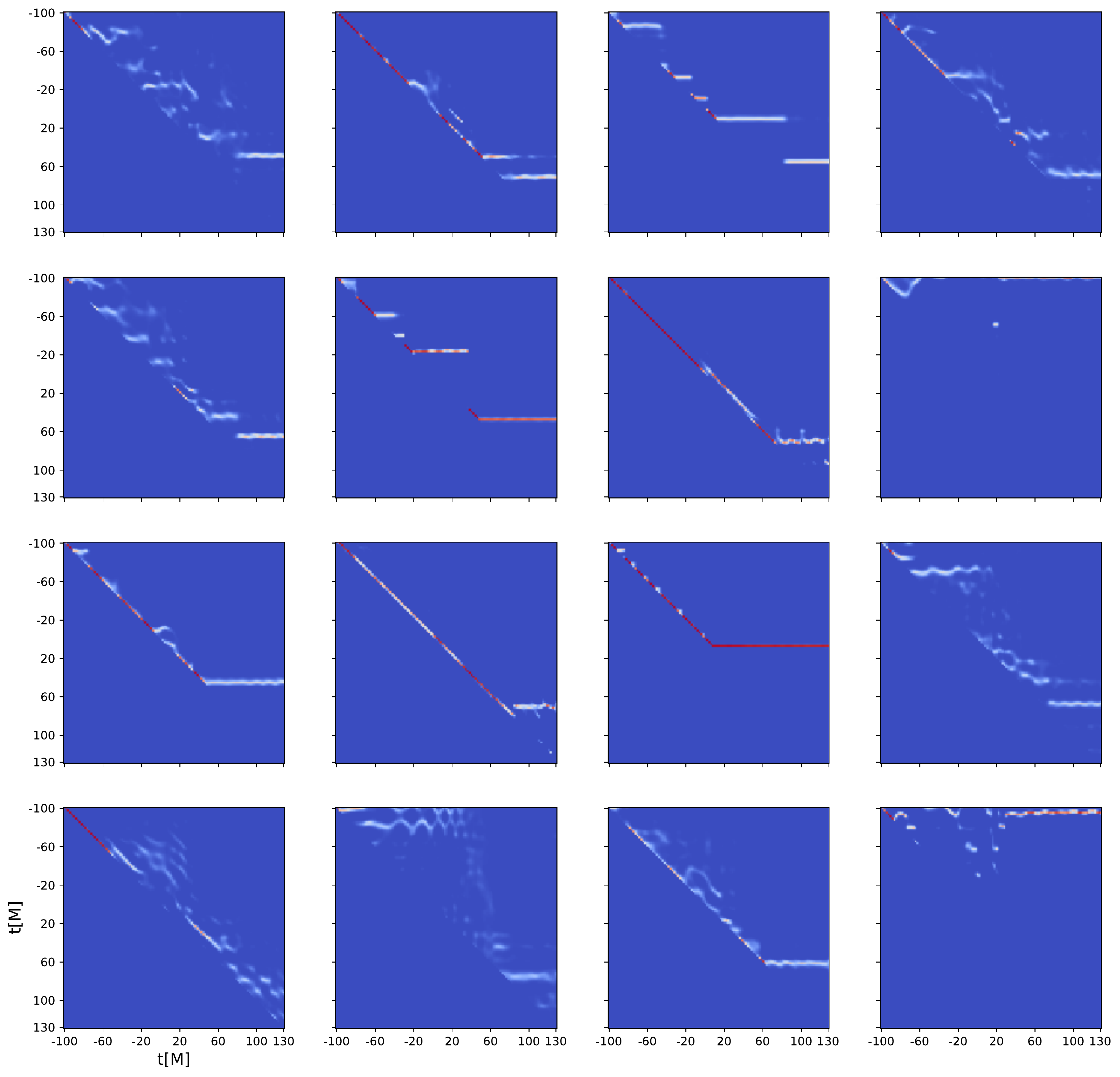}
} 
\caption{As Figure~\ref{fig:all_cross_attention_heads} but 
now for self attention heads.}
\label{fig:all_self_attention_heads}
\end{figure*}


\end{document}